\documentclass[final,1p,times]{elsarticle}

\usepackage{amssymb}
\usepackage{amsmath}
\usepackage{epstopdf}
\usepackage{float}
\usepackage{color,soul}
\usepackage[colorlinks=true,linkcolor=black, citecolor=blue, 
urlcolor=blue]{hyperref}

\journal{Computer Physics Communications}

\begin{document}

\begin{frontmatter}


\title{FORTRESS: FORTRAN programs for solving coupled 
       Gross-Pitaevskii equations for
	spin-orbit coupled spin-1 Bose-Einstein condensate}

\author[label1]{Pardeep Kaur}
\author[label2]{Arko Roy}
\author[label1]{Sandeep Gautam\corref{author}}
\cortext[author] {Corresponding author.\\\textit{E-mail address:} 
pardeepmakkar23@gmail.com, arko.roy@unitn.it, sandeep@iitrpr.ac.in}
\address[label1]{Department of Physics, 
	Indian Institute of Technology Ropar, Rupnagar, Punjab 140001, India}
\address[label2]{INO-CNR BEC Center and Dipartimento di Fisica, 
    Universit{\`a} di Trento, 38123 Trento, Italy}

\begin{abstract}
Here, we present simple and efficient numerical scheme to study static and
dynamic properties of spin-1 Bose-Einstein condensates (BECs) with 
spin-orbit (SO) coupling by solving three coupled Gross-Pitaevskii equations 
(CGPEs) in three-, quasi-two and quasi-one dimensional systems. We provide 
a set of three codes developed in FORTRAN 90/95 programming language with 
user defined '{\em option}' of imaginary and real-time propagation. We present 
the numerical results for energy, chemical potentials, and component densities
for the ground state and compare with the available results from the 
literature. The results are presented for both the ferromagnetic and 
antiferromagnetic spin-1 BECs with and without SO coupling. To improve the 
computational speed, all the codes have the option of OpenMP 
parallelization. We have also presented the results for
speedup and efficiency of OpenMP parallelization for the three codes
with both imaginary and real-time propagation.
\end{abstract}



\end{frontmatter}

{\bf PROGRAM SUMMARY}

\begin{small}
\noindent
	{\em Program Title:} FORTRESS                                          \\
	{\em Licensing provisions:} MIT\\
	{\em Programming language:} (OpenMP) FORTRAN 90/95\\
	{\em Computer:} Intel(R) Xeon(R) Platinum 8180 CPU @ 2.50GHz\\                            
{\em Operating system}: General\\
{\em RAM}: Will depend on array sizes.\\
{\em Number of processors used}: (OPENMP\_THREADS used) 1 for serial and 8 with OpenMP in case of 1D code; 
8 processors for 2D code, 16 processors for 3D code\\
{\em External routines/libraries:} FFTW 3.3.8\\
	{\em Journal reference of previous version:} None                  \\
	{\em Nature of problem:} To solve the coupled Gross-Pitaevskii equations for 
spin-1 BEC with anisotropic spin-orbit coupling using the time-splitting 
spectral method.\\
	{\em Solution method:}
We use the time-splitting Fourier spectral method to solve the coupled
Gross-Pitaevskii equations. The resulting equations are evolved in
imaginary time to obtain the ground state of the system or in real-time
to study the dynamics.\\
\end{small}

\section{Introduction}
Over the past few decades, the study on cold dilute atomic gases has grown 
immensely since the experimental realization of Bose-Einstein condensation 
of bosonic gases in 1995 \cite{anderson1995observation}, a remarkable 
milestone in the field of ultracold atoms. In these early experiments 
\cite{anderson1995observation}, magnetic traps were used giving rise to scalar 
Bose-Einstein condensate (BEC) having frozen spin degrees of freedom. Optical 
traps on the other hand can trap all the hyperfine spin states of spin-$f$ 
ultracold bosonic gas with $f$ as the total spin per atom 
\cite{stamper1998optical}. The advent of these optical traps led to the 
experimental realization of $2f+1$ component Bose-Einstein condensates (BECs),
corresponding to spin projection quantum number $m_f = -f,-f+1,\ldots +f$, 
and is known as spinor-BECs having $f = 1, 2$ and 3
\cite{stenger1998spin1}. Unlike most of the solid-state 
materials, in which spin-orbit (SO) coupling originates due to the 
relativistic effects, there was no spin-orbit coupling in the spinor 
BECs in this early set of experiments \cite{stenger1998spin1}. 
However, SO coupling can be engineered in spinor BECs by controlling the 
atom-light interaction leading to the generation of artificial non-Abelian 
Gauge potentials coupled to the atoms \cite{osterloh2005cold}. SO coupling 
was first engineered in a BEC of $^{87}$Rb \cite{lin2011spin} by dressing two 
of its internal spin states from within the ground electric manifold 
($5S_{1/2}, f=1)$ with a pair of lasers giving rise to equal strengths of 
Rashba \cite{bychkov1984oscillatory} and Dresselhaus 
\cite{dresselhaus1955spin} terms which has attracted a lot of interest on
experimental \cite{aidelsburger2011experimental} and theoretical fronts 
\cite{wang2010spin, zhai2012spin,sinha2011trapped} .  
SO coupling plays a key role in exotic phenomenon like spin-Hall effect 
\cite{galitski2013spin}, topological insulators \cite{hasan2010colloquium} 
and  has motivated new developments in spintronic devices 
\cite{koralek2009emergence}, hybrid structures \cite{avsar2014spin}, and 
topological quantum computation \cite{sau2010generic}, etc. Being highly 
tunable system and offering an unprecedented level of control, 
SO-coupled BEC has become an ideal quantum simulator to study these 
fascinating SO-coupled systems. More recently, SO coupling has been realized 
experimentally in spin-1 $^{87}$Rb \cite{campbell2016magnetic} which has 
stimulated more theoretical \cite{liu2019skyrmions} and experimental 
\cite{li2017stripe} investigations. 
In the domain of strongly correlated electronic systems, recently, SO 
coupling has been employed to drive metal-insulator transition
\cite{zwartsenberg20}. 

To describe a spin-1 BEC, the mean-field theory was developed independently by 
Ho \cite{ho1998spinor} and Ohmi {\em et al.} \cite{ohmi1998bose}. In 
mean-field approximation, an SO-coupled spin-1 BEC is described by a set of 
three coupled time dependent nonlinear partial differential equations with 
first order derivative in time and first and second order derivatives in 
space \cite{wang2010spin}. Since there is no general analytic approach to 
solve a set of coupled Gross-Pitaevskii equations (CGPEs), one needs to solve 
the equations numerically, and this has spurred many studies on the numerical 
solutions of spin-1 BEC \cite{wang2007time,bao2008computing,bao2013efficient}.
A wide range of numerical techniques have been employed in literature to study
single component scalar \cite{ruprecht1995time,chiofalo2000ground,bao2004computing,
antoine2013computational,muruganandam2009fortran,bao2003numerical}, multicomponent
 scalar \cite{chang2005gauss} as well as spinor BECs 
\cite{wang2007time,bao2008computing,bao2013efficient,wang2014projection}. 
One of the most widely used method to determine the ground state of a scalar 
BECs is the imaginary time method followed by an appropriate discretization 
scheme to evolve the resultant gradient flow equations 
\cite{chiofalo2000ground,bao2004computing}. The extension of this method to 
compute the ground states of spin-1 BEC is not straightforward, as there are 
only two constraints, i.e. the conservation of total number of atoms 
and longitudinal magnetization, while one would need three projection 
parameters for normalization of three components of wavefunction 
\cite{wang2007time,bao2008computing,bao2013efficient}. However, imaginary 
time method has been used in the literature with the simultaneous 
conservation of norm and magnetization achieved through the introduction of 
the third normalization condition \cite{wang2007time,bao2008computing,
bao2013efficient}. There have been different discretization schemes used 
which include, among others, centered finite difference scheme and spectral methods for 
spatial discretization and forward Euler, backward Euler, and Crank-Nicolson 
schemes for time discretization. The non-linear terms can be handled easily 
by first using the time-splitting technique, which in the case of scalar 
Gross-Pitaevskii (GP) equation amounts to approximating the solution by
successively solving two equations- one of which is just a free particle 
Schr\"odinger equation, and the other containing the non-linear term can be 
solved exactly \cite{antoine2013computational}. The free particle 
Schr\"odinger equation can be handled by Crank-Nicolson 
\cite{muruganandam2009fortran} or spectral discretization 
\cite{bao2003numerical,bao2002time}. In the present work, we use the 
Fourier spectral discretization for solving the free particle Schr\"odinger 
equation. A couple of advantages of choosing this method: firstly it can be 
extended easily to the higher dimensional systems because of the ease of 
dealing with the differential operators in Fourier space, and secondly its 
spectral accuracy. It is worth pointing out here that even with growing number of 
experimental and theoretical investigations related to SO-coupled spin-1 BECs, 
the technical details of dealing with SOC terms from the numerical
 point of view is still lacking in the literature. This sets the motivation 
for this current work. There are numerous software packages written in 
different programming languages (C, C++, FORTRAN, MATLAB, etc.) to deal with 
solving single or two-coupled GP equations, with and without dipolar interactions, 
under rotating traps, etc. \cite{kumar_19}. 
However, to the best of our knowledge, packages dealing with SO coupled spin-1 BEC 
are unavailable. We make here an attempt to bridge this gap, and make our codes general 
enough to include (a) anisotropic SO coupling, (b) explicit Rashba or 
Dresselhaus type of coupling, or (c) mixture of both.
Each of the three codes has the option of 
{\em imaginary-time} and {\em real-time} propagation to be chosen
by the user. The purpose of imaginary-time propagation, which is also 
referred to as normalized gradient flow method \cite{bao2008computing,
bao2013efficient}, is to find the stationary state solutions of the system, 
whereas the real-time propagation allows the user to study the real-time 
dynamics. We use imaginary time propagation to find the ground state solutions
of SO-coupled quasi-one-dimensional (q1D), quasi-two-dimensional (q2D) and 
three-dimensional (3D) spin-1 BECs.

We use the time-splitting technique \cite{wang2007time,
antoine2013computational,muruganandam2009fortran} to split the CGPEs into 
four sets of equations where each set (consisting of three equations) is 
amenable to be numerically solvable by an appropriate method. These four sets 
of equations are solved successively as per the standard Lie-splitting 
prescription, which is first order accurate in time for two non-commuting 
operators. If the solution of the CGPEs is known at time $t$, say $\Phi(t)$, 
then Lie splitting approximates the solution of the CGPEs at time $t+\delta t$
with the solution obtained by successively solving the aforementioned four 
sets of equations, wherein the solution to each set serves as the initial 
(transient) solution for the following set; except for the first set of 
equations, whose initial solution is $\Phi(t)$. We term the method described 
above as time-splitting real-time propagation \cite{wang2007time,
antoine2013computational,muruganandam2009fortran}. To calculate the ground 
state solutions, we use imaginary-time propagation \cite{wang2007time,
bao2008computing,bao2013efficient} which takes any initial guess to the 
ground state wavefunction after sufficiently large number of time steps; 
as is expected, the number of time steps needed to obtain a converged ground 
state solution depends crucially on the initial guess.  

The main focus of the present paper is to provide efficient and easy to 
implement numerical scheme to solve the CGPEs with anisotropic SO coupling 
\cite{wang2010spin} in imaginary time or real-time. We have implemented the 
numerical scheme via a set of FORTRAN 90/95 codes which can be easily used by 
the students and the researchers working on SO coupled spin-1 BECs. We have 
used harmonic potentials for trapped systems which is widely used in 
experiments, nonetheless the use of Fourier spectral technique makes the 
codes ideal to study the homogeneous system, which is pertinent in the context
of the experimental realization of the box trapping potential 
\cite{ville_18}. 
We present the results for energy, chemical potentials and densities of 
ground state wave functions obtained with the codes and compare them with the 
earlier reported results in the literature \cite{bao2008computing,
bao2013efficient}.

The paper is organized as follows. In section \ref{section-II}, we describe 
the mean-field CGPEs with SO coupling for spin-1 condensate, and then 
the dimensionless formulation of these equations in three dimension. This is 
followed by the reduction of the set for q2D and q1D BECs. In section 
\ref{section-III}, we discuss the details of the numerical approach to solve 
these equations in one dimension, followed by the discussion on q2D and 3D 
spinor BECs. For the sake of brevity, in q2D and 3D cases, the emphasis of 
the discussion is on the additional changes to the q1D scheme. We conclude 
the section with a discussion on discretization scheme in real and Fourier 
space for sake of completeness. In section \ref{section-IV}, we present the 
description of FORTRAN programs which include definition of the various data 
variables or constant parameters and the functions of the various subroutines. 
In section \ref{section-V}, we present the results for performance 
parameters like speedup and efficiency of OpenMP programs for both imaginary 
and real-time propagations. 
In section \ref{section-VI}, we present the results for energy, chemical 
potentials, component wave-functions or densities and 
compare them with the ones reported by other researchers.

\section{Coupled Gross-Pitaevskii equations for spin-orbit coupled BEC}	
\label{section-II}
In 3D case, the single particle Hamiltonian of spin-1 BEC in the presence 
of anisotropic \cite{bychkov1984oscillatory,campbell2016rashba} SO 
coupling is given by \cite{wang2010spin,zhai2012spin}

\begin{equation}
H_{0} = \frac{p_x^2+p_y^2+ p_z^2}{2m} + \gamma_x p_x \Sigma_x+\gamma_y p_y 
\Sigma_y + \gamma_z p_z \Sigma_z  ,
\end{equation}
where $p_x = -i\hbar\partial/\partial x$, $p_y = -i\hbar\partial/\partial y$, 
and $p_z = -i\hbar\partial/\partial z$ correspond to the momentum operators 
along $x$, $y$ and $z$ directions, respectively. Also, $m$ is the mass of each 
atom and $\Sigma_x$, $\Sigma_y$, and $\Sigma_z$ are the irreducible matrix 
representations of the $x$, $y$ and $z$ components of the spin-1 angular 
momentum operator, respectively, which are given by

\begin{eqnarray}
\Sigma_x&=\frac{1}{\sqrt 2} \begin{pmatrix}
0 & 1 & 0 \\
1 & 0  & 1\\
0 & 1 & 0
\end{pmatrix}, \quad
\Sigma_y=\frac{1}{\sqrt 2 i} \begin{pmatrix}
0 & 1 & 0 \\
-1 & 0  & 1\\
0 & -1 & 0
\end{pmatrix},  \quad
\Sigma_z=\begin{pmatrix}
1 & 0 & 0 \\
0 & 0  & 0\\
0 & 0 & -1
\end{pmatrix},\label{spin_matrices}
\end{eqnarray}
where $\gamma_x$, $\gamma_y$, and $\gamma_z$ are the strengths of SO coupling. 
In standard isotropic SO coupling, $\gamma_x = \gamma_y = \gamma_z 
= \hbar k_{r}/m$ realized by using two counter-propagating  Raman lasers of 
wavelength $\lambda_{r}$ aligned at an angle ${\beta_{r}}$ and $k_{r}$ is 
given by $k_{r} =(2 \pi \sin{ \beta_r/2})/ \lambda_{r}$.

For weakly interacting SO-coupled spin-1 BEC, the properties of system are 
well described under mean-field approximation by the following coupled 
Gross-Pitaevskii equations (CGPEs) 
\cite{wang2010spin,ho1998spinor,ohmi1998bose,wang2007time}
\begin{subequations}
\begin{eqnarray}
i\hbar\frac{\partial \psi_1}{\partial t} &=& \mathcal{H}
\psi_1 +c_2(\rho_0 + \rho_{-})\psi_1+c_2\psi_{-1}^*\psi_0^2 
- \frac{i\hbar}{\sqrt{2}}\left(\gamma_x\frac{\partial\psi_0}{\partial x}
-i\gamma_y\frac{\partial\psi_0}{\partial y}
+\sqrt{2}\gamma_z\frac{\partial\psi_1}{\partial z}\right) ,\label{cgpe-1}\\
i\hbar\frac{\partial \psi_0}{\partial t} &=& \mathcal{H}
\psi_0 +c_2\rho_{+} \psi_0+ 2 c_2\psi_1\psi_{-1}\psi_0^* - \frac{i\hbar}
{\sqrt{2}}\left[\gamma_x\left(\frac{\partial\psi_1}{\partial x}
+\frac{\partial\psi_{-1}}{\partial x}\right)+
i\gamma_y \left(\frac{\partial\psi_1}{\partial y}
-\frac{\partial\psi_{-1}}{\partial y}\right)\right] ,\label{cgpe-2}\\
i\hbar\frac{\partial \psi_{-1}}{\partial t} &=&
\mathcal{H}\psi_{-1} +c_2(\rho_0-\rho_{-}  )\psi_{-1}+c_2\psi_1^*\psi_0^2 
- \frac{i\hbar}{\sqrt{2}}\left(\gamma_x\frac{\partial\psi_0}{\partial x}
+i\gamma_y\frac{\partial\psi_0}{\partial y}
-\sqrt{2}\gamma_z\frac{\partial\psi_{-1}}
{\partial z}\right) ,\label{cgpe-3}
\end{eqnarray}
\end{subequations}
where
\begin{equation}
\mathcal{H}=(-\frac{\hbar^2}{2m} \mathbf{\nabla^2}
+V(\mathbf{x})+c_0\rho),\quad  \mathbf{x} = (x,y,z),
\end{equation}
and $\Psi = (\psi_1(\mathbf{x},t),\psi_0(\mathbf{x},t),
\psi_{-1}(\mathbf{x},t))^T$ with $\psi_1$, $\psi_0$ and $\psi_{-1}$ as the 
component wavefunctions, and $V(\mathbf{x})= 
m(\omega_x^2x^2+\omega_y^2y^2
+\omega_z^2z^2)/2$ is 3D harmonic trap. 
Also,
\begin{equation}
c_0 = \frac{4\pi\hbar^2(a_0+2a_2)}{3m},
\quad c_2 =\frac{4\pi\hbar^2(a_2-a_0)}{3m},   
\quad \nabla^2 = \frac{\partial^2}{\partial x^2}
+\frac{\partial^2}{\partial y^2}+\frac{\partial^2}{\partial z^2},
\end{equation}
where $a_0$ and $a_2$ correspond to the $s$-wave scattering lengths in total 
spin $0$ and $2$ channels, respectively; $\omega_x$, $\omega_y$ and $
\omega_z$ are the confining trap frequencies along $x$, $y$ and $z$ directions,
respectively; $\rho_{\pm} = \rho_{+1}\pm\rho_{-1}$ where  $\rho_j = |\psi_j|^2$
with $j=1,0,-1$ are the component densities and 
$\rho = \sum_{j=-1}^1|\psi_j|^2$ is the total density. 

\subsection{Important conserved quantities of Spin-1 BEC}
Three important conserved quantities of spin-1 BEC are total number of 
particles $N$, longitudinal magnetization $\cal M$ (which is conserved on 
the time scale of spin-1 BEC experiments), and total energy $E$. These are 
given as 
\begin{subequations}
\begin{eqnarray}
N &=& \int_{-\infty}^\infty d\mathbf{x} \sum_{j=-1}^1|\psi_j(\mathbf{x})|^2,\\
{\cal M} &=& \sum_{j=-1}^{1} \int j |\psi_j({\bf x},t)|^2 d{\bf x},
\end{eqnarray}
\end{subequations}
\begin{eqnarray}
E &=& \int d\textbf{x} \Biggl[\sum_{j=-1}^1 \psi_j^* \left(-\frac{\hbar^2 
\nabla^2}{2 m}+V\right) \psi_j + \frac{c_0}{2} \rho^2 + \frac{c_2}{2}(\rho_1 
+ \rho_0 - \rho_{-1}) \rho_1 +\frac{c_2}{2}(\rho_0+\rho_{-1}-\rho_1)\rho_{-1}
\nonumber\\
&& +c_2(\psi_{-1}^*{\psi_0}^2\psi_1^* +\psi_{-1}{{\psi_0}^2}^*\psi_1) 
 +\frac{c_2}{2} (\rho_1+\rho_{-1})\rho_0-\frac{i \hbar \gamma_x}{\sqrt{2}} 
{\psi_0}^*\left(\frac{\partial\psi_1}{\partial{x}}+ 
\frac{\partial\psi_{-1}}{\partial{x}}\right)
\nonumber\\
&& +\frac{\hbar}{\sqrt{2}} 
{\psi_0}^*\left(\gamma_y\frac{\partial\psi_1}{\partial{y}}
- \gamma_x\frac{\partial\psi_{-1}}
{\partial{x}}\right) -\frac{i \hbar \gamma_x}{\sqrt{2}}({\psi_1}^* 
+ {\psi_{-1}}^*)
\frac{\partial\psi_0}{\partial{x}}-\frac{\hbar \gamma_y}{\sqrt{2}}({\psi_1}^* 
-{\psi_{-1}}^*)\frac{\partial\psi_0}{\partial{y}}\nonumber\\
&&-i \hbar \gamma_z \left({\psi_1}^*\frac{\partial\psi_1}{\partial{z}}- 
{\psi_{-1}}^*\frac{\partial\psi_{-1}}{\partial{z}}\right) \Biggr].
\end{eqnarray}
\subsection{Chemical potential}
For stationary states, the wavefunctions have the trivial time dependence 
$\psi_j(\mathbf x,t) = e^{-i \mu_j t} \psi_j({\bf x})$ through the Madelung 
transformation. By plugging this into Eqs. (\ref{cgpe-1})-(\ref{cgpe-3}), 
the time independent CGPEs are
\begin{subequations}
\begin{eqnarray}
\mu_{1} \psi_1 &=& [\mathcal{H} + c_2 (\rho_0 + \rho_{-})]\psi_1 
+c_2 {\psi_{-1}}^* {\psi_0}^2- \frac{i\hbar}{\sqrt{2}}
\left(\gamma_x\frac{\partial\psi_0}{\partial x}
-i\gamma_y\frac{\partial\psi_0}{\partial y}
+\sqrt{2}\gamma_z\frac{\partial\psi_1}{\partial z}\right),\\ 
\mu_{0} \psi_0 &=& \left[\mathcal{H} + c_2 \rho_{+}  \right]\psi_0 
+ 2 c_2 {\psi_{0}}^* {\psi_1} {\psi_{-1}}- \frac{i\hbar}{\sqrt{2}}
\left(\gamma_x\frac{\partial\psi_1}{\partial x}
+i\gamma_y\frac{\partial\psi_1}{\partial y}
+\gamma_x\frac{\partial\psi_{-1}}{\partial x}
-i\gamma_y\frac{\partial\psi_{-1}}{\partial y}\right),\\
\mu_{-1} \psi_{-1} &=& \left[\mathcal{H} + c_2 (\rho_0 - \rho_{-}) \right]
\psi_1 +c_2 {\psi_{1}}^* {\psi_0}^2- \frac{i\hbar}{\sqrt{2}}
\left(\gamma_x\frac{\partial\psi_0}{\partial x}
+i\gamma_y\frac{\partial\psi_0}{\partial y}
-\sqrt{2}\gamma_z\frac{\partial\psi_{-1}}{\partial z}\right),
\end{eqnarray}
\end{subequations}
where $\mu_{1}, \mu_{0}$ and $\mu_{-1}$ are the chemical potentials of the 
three components. These equations can be used to define the chemical 
potential functionals analogous to energy functional.

\subsection{Dimensionless formulation of 3D CGPEs}
Eqs. (\ref{cgpe-1}) - (\ref{cgpe-3}) can be transformed into dimensionless
form by introducing the following dimensionless variables
\begin{equation}
\tilde{t} = 2 \omega_x t,\quad
~\tilde{\mathbf{x}} = \frac{\mathbf{x}}{a_{\rm osc}},\quad
~\phi_j(\tilde{\mathbf{x}},\tilde{t}) = 
\frac{{a_{\rm osc}}^{3/2}}{\sqrt{N}}\psi_j(\tilde{\mathbf{x}},\tilde{t}),\quad
a_{\rm osc}=\sqrt{\frac{\hbar}{m\omega_{x}}}      
\end{equation}
where $a_{\rm osc}$ is the oscillator length. This basically fixes the units 
of length, time, density, and energy as $a_{\rm osc}$, $1/2\omega_{x}$, 
$a_{\rm osc}^{-3}$, and  $\hbar\omega_{x}$, respectively. After substitution of 
these new parameters and removing all tildes for notational simplicity, we get 
the following dimensionless CGPEs in 3D \cite{wang2007time,gautam2018three} 
\begin{subequations}
\begin{eqnarray}
i\frac{\partial \phi_1}{\partial t} &=&
\mathcal{H}\phi_1 +2c_2(\rho_0  +\rho_{-})\phi_1+2c_2\phi_{-1}^*\phi_0^2 - 
\sqrt{2}i\left(\gamma_x\frac{\partial\phi_0}{\partial x}
-i\gamma_y\frac{\partial\phi_0}{\partial y}+\sqrt{2}\gamma_z
\frac{\partial\phi_1}
{\partial z}\right),\label{cgpet-1}\\
i\frac{\partial \phi_0}{\partial t} &=&
\mathcal{H}\phi_0 +2c_2\rho_{+}\phi_0+ 4 c_2\phi_1\phi_{-1}\phi_0^* 
- \sqrt{2}{i}\left(\gamma_x\frac{\partial\phi_1}{\partial x}
+i\gamma_y\frac{\partial\phi_1}{\partial y}
+\gamma_x\frac{\partial\phi_{-1}}{\partial x}
-i\gamma_y\frac{\partial\phi_{-1}}{\partial y}\right),\label{cgpet-2}\\
i\frac{\partial \phi_{-1}}{\partial t} &=&
\mathcal{H}\phi_{-1} +2c_2(\rho_0-\rho_{-}  )\phi_{-1}+2c_2\phi_1^*\phi_0^2 
- \sqrt{2}i\left(\gamma_x\frac{\partial\phi_0}{\partial x}
+i\gamma_y\frac{\partial\phi_0}{\partial y}
-\sqrt{2}\gamma_z\frac{\partial\phi_{-1}}
{\partial z}\right),\label{cgpet-3}
\end{eqnarray}
\end{subequations}
where  
\begin{eqnarray}
\mathcal{H} &=- \mathbf{\nabla^2} +2V(\mathbf{x})+2c_0\rho ,\quad V({\bf x}) 
&= ({\alpha_x}^2 x^2 +{\alpha_y}^2 y^2+{\alpha_z}^2 z^2)/2,
\label{trap3D}
\end{eqnarray}  
$\alpha_\eta = \omega_\eta/\omega_{x}$ with
$\eta = x, y, z$ and new $c_0$, $c_2$ and $\gamma$ are given by
\begin{equation}
c_0 = \frac{4\pi N(a_0+2a_2)}{3 a_{\rm osc}},\quad
c_2 =\frac{4\pi N(a_2-a_0)}{3{a_{\rm osc}}},\quad \gamma_x = \gamma_y = \gamma_z =  k_r a_{\rm osc}.   
\label{interaction3d}
\end{equation}
Also, $\rho_j = |\phi_j|^2$ with $j=1,0,-1$ are the component densities,
$\rho = \sum_{j=-1}^1|\phi_j|^2$ is the total density, and now it is normalized
to unity, i.e. $\int \rho d{\mathbf x} = 1$.

\subsection{CGPEs for q2D Spin-1 BEC}\label{CGPEsin2D}
If the trapping frequencies along any direction, let us say $z$ is much larger 
than the geometric mean of frequencies along other two directions, i.e $x$ and
$y$, then $\alpha_x = 1$, $\alpha_y \approx 1$ and $\alpha_z \gg \alpha_x$ 
\cite{wang2007time}. In this case, the dimensionless generalized CGPEs in 3D 
can be approximated by 2D equations by choosing \cite{salasnich2002effective}
\begin{equation}
\phi_j(x,y,z,t) = \phi_j(x,y,t)\phi_{\rm ho}(z), \quad
\phi_{\rm ho}(z) = ({\alpha_z}/\pi)^{1/4} \exp\left({-\alpha_z z^2/2}\right).
\end{equation}
Generalized dimensionless CGPEs in 2D are given by 
\cite{wang2007time,gautam2017vortex}
\begin{subequations}
\begin{eqnarray}
i\frac{\partial \phi_1}{\partial t} &=&
\mathcal{H}\phi_1 +2{c_2}(\rho_0  +\rho_{-})\phi_1+2{c_2}\phi_{-1}^*\phi_0^2 
- \sqrt{2}i\left(\gamma_x\frac{\partial\phi_0}{\partial x}
-i\gamma_y\frac{\partial\phi_0}{\partial y}\right),\label{cgpet2d-1}\\
i\frac{\partial \phi_0}{\partial t} &=&
\mathcal{H}\phi_0 +2{c_2}\rho_{+}\phi_0+ 4{c_2}\phi_1\phi_{-1}\phi_0^* - 
\sqrt{2}{i}\left(\gamma_x\frac{\partial\phi_1}{\partial x}
+i\gamma_y\frac{\partial\phi_1}{\partial y}
+\gamma_x\frac{\partial\phi_{-1}}{\partial x}
-i\gamma_y\frac{\partial\phi_{-1}}{\partial y}\right),\label{cgpet2d-2}\\
i\frac{\partial \phi_{-1}}{\partial t} &=&
\mathcal{H}\phi_{-1} +2{c_2}(\rho_0-\rho_{-}  )\phi_{-1}
+2{c_2}\phi_1^*\phi_0^2 - \sqrt{2}i\left(\gamma_x\frac{\partial\phi_0}
{\partial x}+i\gamma_y\frac{\partial\phi_0}{\partial y}\right),\label{cgpet2d-3}
\end{eqnarray}
\end{subequations}
where
\begin{equation} 
\mathcal{H}=- \nabla_{xy}^2+2V(\mathbf{x})+2 {c_0}\rho,\quad 
\nabla_{xy}^2 = 
\frac{\partial^2}{\partial x^2}+\frac{\partial^2}{\partial y^2},
\quad 
\mathbf{x} = (x,y). 
\end{equation}
The trapping potential $V(\mathbf{x})$ and interaction parameters $c_0$ and
$c_2$ are now defined as
\begin{equation}
V(\mathbf{x})= \frac{1}{2}({\alpha_x}^2 x^2 +{\alpha_y}^2 y^2),\quad 
c_0 = \sqrt{\frac{{\alpha_z}}{2\pi}}\frac{4\pi N(a_0+2a_2)}{3{a_{\rm osc}}},
\quad 
c_2 = \sqrt{\frac{{\alpha_z}}{2\pi}}\frac{4\pi N(a_2-a_0)}{3{a_{\rm osc}}}.
\label{interaction2d}
\end{equation} 

\subsection{CGPEs for q1D Spin-1 BEC}
If the trap is much stronger along two directions, say $y$ and $z$ compared 
to the $x$ direction then $\alpha_x = 1, \alpha_y \gg \alpha_x, 
\alpha_z \gg \alpha_x$ then by assuming \cite{salasnich2002effective} 
\begin{equation}
\phi_j(x,y,z,t) = \phi_j(x,t)\phi_{\rm ho}(y,z), \quad
\phi_{\rm ho}(y,z) = ({\sqrt{\alpha_y \alpha_z}/\pi})^{1/2}
\exp[-(\alpha_y y^2 + \alpha_z z^2) /2],
\end{equation}
Eqs. (\ref{cgpet-1})-(\ref{cgpet-3}) can be reduced into quasi-1D 
equations \cite{wang2007time,gautam2015mobile}
\begin{subequations}
\begin{eqnarray}
i\frac{\partial \phi_1}{\partial t} &=&
\mathcal{H} \phi_1 +2{c_2}(\rho_0  +\rho_{-})\phi_1+2{c_2}\phi_{-1}^*\phi_0^2 
- \sqrt{2}i\gamma_x \left(\frac{\partial\phi_0}{\partial x}\right),
\label{cgpet1d-1}\\
i\frac{\partial \phi_0}{\partial t} &=&
\mathcal{H} \phi_0 +2{c_2}\rho_{+}\phi_0+ 4 {c_2}\phi_1\phi_{-1}\phi_0^* 
- \sqrt{2}{i\gamma_x}\left(\frac{\partial\phi_1}{\partial x}
+\frac{\partial\phi_{-1}}{\partial x}\right),\label{cgpet1d-2}\\
i\frac{\partial \phi_{-1}}{\partial t} &=&
\mathcal{H} \phi_{-1} +2{c_2}(\rho_0-\rho_{-}  )\phi_{-1}+2{c_2}\phi_1^*
\phi_0^2 - \sqrt{2}i\gamma_x\left(\frac{\partial\phi_0}{\partial x}\right),
\label{cgpet1d-3}
\end{eqnarray}
\end{subequations}
where
\begin{eqnarray*}
\mathcal{H} &=- \frac{\partial^2 }{\partial x^2}
+2V({x})+2{c_0}\rho, \quad V(x) = \frac{1}{2}\gamma_x^2 x^2,\\
{c_0} &= \frac{\sqrt{\alpha_y \alpha_z}}{2\pi}
\frac{4\pi N(a_0+2a_2)}{3{a_{\rm osc}}},\quad
{c_2} = \frac{\sqrt{\alpha_y \alpha_z}}{2\pi}
\frac{4\pi N(a_2-a_0)}{3{a_{\rm osc}}}.
\end{eqnarray*}

\section{Numerical Methods} \label{numerical methods}
\label{section-III}
\subsection{Solution of q1D CGPEs}
Starting with the simplest case of q1D spin-1 BEC, 
Eqs (\ref{cgpet1d-1})-(\ref{cgpet1d-3}) can be written in simplified form as 
\begin{equation}
i\frac{\partial\Phi}{\partial t} = {\rm H}\Phi. \label{cgpe}
\end{equation}
Here, $\Phi=(\phi_{1},\phi_{0},\phi_{-1})^T$ and Hamiltonian ${\rm H}$ 
consists of different terms involving kinetic energy operator $H_{\rm KE}$, 
trapping potential plus terms resulting from spin-preserving collisions 
$H_{\rm SP}$, terms corresponding to spin-exchange collisions $H_{\rm SE}$, and 
spin-orbit coupling $H_{\rm SOC}$. Eq. (\ref{cgpe}) can then be written as
\begin{equation}
i\frac{\partial\Phi}{\partial t} = (H_{\rm KE} + H_{\rm SP} + H_{\rm SE} 
+ H_{\rm SOC})\Phi, \label{split}
\end{equation}
where $H_{\rm KE}$, $H_{\rm SP}$, $H_{\rm SE}$ and $H_{\rm SOC}$ are 
$3\times3$ matrix 
operators defined as
\begin{subequations}
\begin{eqnarray}
H_{\rm KE} &=
\begin{pmatrix}
- \partial_{xx} & 0 &0 \\
0 &-\partial_{xx} &0 \\
0 & 0 & - \partial_{xx}
\end{pmatrix}{\label{KEmatrix}},\quad
H_{\rm SOC} &=-{\sqrt{2}}i\gamma_x \begin{pmatrix}
0 & \partial_{x}&0 \\
\partial_{x}& 0 &\partial_{x}
\\
0 & \partial_{x}  & 0
\end{pmatrix}{\label{SOCmatrix}},
\end{eqnarray}
\begin{equation}
H_{\rm SE} =
\begin{pmatrix}
0 & 2{c_2}\phi_0\phi_{-1}^*&0 \\
2{c_2}\phi_0^*\phi_{-1} & 0 &2{c_2}\phi_0^*\phi_{1} \\
0 & 2{c_2}\phi_0{\phi_1}^* & 0
\end{pmatrix}{\label{SEmatrix}},
\end{equation}
\begin{eqnarray}
H_{\rm SP} =2
\begin{pmatrix}
{V}+{c_0}\rho+{c_2}(\rho_0+\rho_{-}) & 0 &0 \\
0 & {V}+{c_0}\rho +{c_2}\rho_{+}&0 \\
0 & 0 & {V}+{c_0}\rho+{c_2}(\rho_0-\rho_{-})
\end{pmatrix}{\label{SPmatrix}}.\quad
\end{eqnarray}
\end{subequations}
To solve these equations (\ref{split}), we use operator splitting which 
has been extensively used in the numerical solutions of non-linear 
Schr\"odinger equation including GP equation 
\cite{muruganandam2009fortran,kumar2019c} 
and coupled GP equations \cite{wang2007time}. Here, we have used first order 
time splitting known as Lie splitting. 

Solution to Eq. (\ref{split}) after time step $\delta t$ is given as
\begin{equation}
\Phi(t+ \delta t) = \hat {U} \Phi(t)\label{unitary},
\end{equation}
which describes the evolution of the wave function by a unitary propagator 
$\hat {U}$ given as
\begin{equation}
\hat{U} = \exp \left[{-i\delta{t} (H_{\rm SP} + H_{\rm SE} + H_{\rm SOC} 
+ H_{\rm KE})}\right],
\end{equation}
The propagator can be approximated by split operator technique as
\begin{equation}
\hat {U} \approx \exp\left({-i {\delta{t}} H_{\rm SP}} \right) 
\exp\left({-i {\delta{t}} H_{\rm SE}} \right)
\exp\left({-i{\delta{t}} H_{\rm SOC}} \right)
\exp\left({-i {\delta{t}} H_{\rm KE}} \right)\label{unitaryapprox}.
\end{equation}
Using (\ref{unitaryapprox}), Eq. (\ref{unitary}) is equivalent to solving 
following equations successively
\begin{subequations}
\begin{eqnarray}
i\frac{\partial \Phi}{\partial t} &=& H_{\rm KE}\Phi, \label{KEpart}\\
i\frac{\partial \Phi}{\partial t} &=& H_{\rm SOC}\Phi, \label{SOCpart}\\
i\frac{\partial \Phi}{\partial t} &=& H_{\rm SE}\Phi, \label{SEpart}\\
i\frac{\partial \Phi}{\partial t} &=& H_{\rm SP}\Phi. \label{SPpart}
\end{eqnarray}
\end{subequations}
Eq. (\ref{KEpart}) can be written as the following set of decoupled equations
\begin{equation}
i\frac{\partial \phi_{j}(x,t)}{\partial t}  = - \frac{\partial^2 \phi_{j}(x,t)}
{\partial x^2}  , \quad  j= -1,0,1. \label{KEpartmom}
\end{equation}
Solution of Eq. (\ref{KEpartmom}) in Fourier space is given as
\begin{equation}
\hat{\phi}_{j}(k_x,t+\delta t) = \hat{\phi}_{j}(k_x,t) 
\exp(-i k_x^2 \delta t),\label{sol_KEpartmom}
\end{equation}
where $\hat{\phi}_{j}$ is the Fourier transform of $\phi_{j}$ and 
$k_x$ is known as Fourier frequency. Now, $\hat{\phi}_{j}(k_x,t+\delta t)$,
transient wavefunction in Fourier space, is the initial value of wavefunction 
for the Fourier transform of Eq. (\ref{SOCpart}), i.e.,  
\begin{equation}
i\frac{\partial\hat{\Phi}(k_x, t)}{\partial t} = 
\hat{H}_{\rm SOC}\hat{\Phi}(k_x,t) \label{SOCpartF}.
\end{equation}
Here $\hat{H}_{\rm SOC}$ is given as
\begin{equation}
\hat{H}_{\rm SOC} = -{\sqrt{2}}i\gamma_x \begin{pmatrix} 
0 & ik_x& 0\\
ik_x  &0 & ik_x \\
0 &  ik_x & 0 
\label{HSOC} 
\end{pmatrix},
\end{equation}
and $\hat{\Phi}(k_x,t)$ is the Fourier transforms of $\Phi(x,t)$.
The solution of equation (\ref{SOCpartF}) is given as \cite{gautam2018three,gautam2017vortex}
\begin{equation}
\begin{split}
\hat{\Phi}(k_x, t+\delta t)&= e^{-i\hat{H}_{\rm SOC}
	\delta t} \hat{\Phi}(k_x, t) = e^{-i\hat{G}} \hat{\Phi}(k_x, t)\\
&= \left(I + \frac{\cos{\beta }-1}{\beta^2}\hat{G}^2-
i\frac{\sin{\beta}}{\beta}\hat{G}\right)\hat{\Phi}(k_x, t),\label{sol_to_soc}
\end{split}
\end{equation}
where $\beta = \sqrt{2}A\delta t$ with $A = {\sqrt{2}}\gamma_x k_x $
and $\hat{G}$ is defined as
\begin{equation}
\hat{G} = \delta t\begin{pmatrix}
0 & A & 0\\
A & 0 & A\\
0 & A & 0
\end{pmatrix}.
\end{equation}
Wavefunction in Eq. (\ref{sol_to_soc}) is in Fourier space and is inverse 
Fourier transformed to obtain the transient wavefunction in co-ordinate space
which serves as the initial solution for Eq. (\ref{SEpart}).
The solution of Eq. (\ref{SEpart}) is now given by
\begin{equation}
\begin{split}
{\Phi}(x, t+\delta t) = e^{-i{H}_{\rm SE}
	\delta t} {\Phi}(x,t) = e^{-i{\hat O}} {\Phi}(x,t)\\
= \left(I + \frac{\cos{\Omega }-1}{\Omega^2}\hat{O}^2-
i\frac{\sin{\Omega}}{\Omega}\hat{O}\right){\Phi}(x,t)
\label{HSE_imp}
\end{split}
\end{equation}
where $H_{\rm SE}$ is given in Eq. (\ref{SEmatrix}) and $\hat{O}$ is defined as
\begin{equation}
\hat{O} = \delta t\begin{pmatrix}
0  & A & 0\\
A^* & 0 & B^*\\
0  & B & 0
\end{pmatrix}.
\end{equation}
with $\Omega = \delta t\sqrt{{|A|^2 +|B|^2}}$,\quad
$A =2{c_2}\phi_0 {\phi_{-1}}^* $ and $ B =2{c_2} \phi_0 {\phi_1}^*  $ \\
The transient wave function we get from here is in configuration space and
is used as an input wavefunction for remaining Eq. (\ref{SPpart}). 
$H_{SP}$ being diagonal, the solution to Eq. (\ref{SPpart}) can be 
calculated analytically as 
\begin{equation}
\Phi(x,t+ \delta t) = \exp\left({-i {\delta{t}} H_{\rm SP}} \right)\Phi(x,t).
\label{HSP_imp}
\end{equation} 
This final wavefunction is solution of Eq. (\ref{unitary}) after 
time $\delta t$.

\subsubsection{Solution of q2D CGPEs}
\label{q2D_section}
The method discussed in previous subsection can be extended to q2D and 3D 
systems with some modifications which we will elaborate in the rest of
this section. In q2D spin-1 BECs, Eqs. (\ref{cgpet2d-1})-(\ref{cgpet2d-3}) 
can again be written in simplified form as Eq. (\ref{cgpe}). 
Here too ${\rm H}$ can be considered as consisting of, aptly defined, 
$H_{\rm KE}, H_{\rm SOC}, H_{\rm SE}$ and $H_{\rm SP}$. Now, $H_{\rm KE}$ 
and $H_{\rm SOC}$ for q2D SO-coupled BECs are given as
\begin{subequations}
\begin{equation}
H_{\rm KE} =
\begin{pmatrix}
- \nabla_{xy}^2 & 0 &0 \\
0 &- \nabla_{xy}^2&0 \\
0 & 0 & - \nabla_{xy}^2
\end{pmatrix}\label{KE2matrix},
\end{equation}
\begin{equation}
H_{\rm SOC} =-{\sqrt{2}}i \begin{pmatrix}
0 &\gamma_x\partial_{x}-i\gamma_y\partial_{y}&0 \\
\gamma_x\partial_{x}+i\gamma_y\partial_{y} & 0 & 
\gamma_x\partial_{x}-i\gamma_y\partial_{y}
\\
0 & \gamma_x\partial_{x}+i\gamma_y\partial_{y}  & 0
\end{pmatrix}{\label{SOC2matrix}},\\
\end{equation}
\end{subequations}
whereas $H_{\rm SE}$ and $H_{\rm SP}$ are 
again defined by Eqs. (\ref{SEmatrix}) and 
(\ref{SPmatrix}), respectively, where $V$, $c_{0}$ and $c_2$ are now given by 
Eq. (\ref{interaction2d}). Again as in q1D systems, solution of 
Eqs. (\ref{cgpet2d-1})-(\ref{cgpet2d-3}) is approximated by solving 
Eqs. (\ref{KEpart})-(\ref{SPpart}) successively.
Solution to  Eq. (\ref{KEpart}) with $H_{\rm KE}$ defined by 
Eq. (\ref{KE2matrix})
in this case is given in Fourier space as
\begin{equation}
\hat{\phi}_{j}(k_x,k_y,t+\delta t) = 
\hat{\phi}_{j}(k_x,k_y,t) \exp[-i (k_x^2+k_y^2) \delta t].
\end{equation}
Fourier transform of Eq. (\ref{SOCpart}) corresponding to 
$H_{\rm SOC}$ given by Eq.  (\ref{SOC2matrix}) is given as 
\begin{equation}
\frac{i\partial\hat{\Phi}(k_x,k_y, t)}{\partial t} = 
\hat{H}_{\rm SOC}\hat{\Phi}(k_x,k_y,t),\label{soc2d}
\end{equation}
where $\hat{H}_{\rm SOC}$ in Fourier space is given as
\begin{equation}
\hat{H}_{\rm SOC} = -{\sqrt{2}}i \begin{pmatrix} 
0 & i\gamma_xk_x+\gamma_yk_y& 0\\
i\gamma_xk_x-\gamma_yk_y  &0 & i\gamma_xk_x+\gamma_yk_y \\
0 &  i\gamma_xk_x-\gamma_yk_y & 0  
\end{pmatrix}.
\end{equation}
Solution to (\ref{soc2d}) is given as \cite{gautam2018three,gautam2017vortex}
\begin{equation}
\begin{split}
\hat{\Phi}(k_x, k_y, t+\delta t)= e^{-i\hat{H}_{\rm SOC}
	\delta t} \hat{\Phi}(k_x ,k_y, t) = e^{-i\hat{G}} 
\hat{\Phi}(k_x, k_y,t)\\
= \left(I + \frac{\cos{\beta }-1}{\beta^2}\hat{G}^2-
i\frac{\sin{\beta}}{\beta}\hat{G}\right)
\hat{\Phi}(k_x,k_y, t),\label{sol_to_soc2d}
\end{split}
\end{equation}
where $\beta = \sqrt{2}|A|\delta t $,\quad
$A = {\sqrt{2}} (\gamma_xk_x-i \gamma_yk_y) $,\quad
$A^* = {\sqrt{2}} (\gamma_xk_x+i \gamma_yk_y) $
and $\hat{G}$ is
defined as
\begin{equation}
\hat{G} = \delta t\begin {pmatrix}
0 & A & 0\\
A^* & 0 & A\\
0 & A^* & 0
\end{pmatrix}.
\end{equation}
Eqs. (\ref{SEpart}) and (\ref{SPpart}) are solved similarly as in q1D case.

\subsubsection{Solutions of 3D CGPEs}
\label{3D_section}
In 3D case too, forms of $H_{\rm SE}$ and $H_{\rm SP}$ are same as defined in 
(\ref{SEmatrix}) and (\ref{SPmatrix}) where $V({\bf x})$, $c_{0}$ and 
$c_2$ are defined in Eqs. (\ref{trap3D})-(\ref{interaction3d}) allowing
us to use the methods discussed in q1D case to solve 
Eqs. (\ref{SEpart})-(\ref{SPpart}). On the other hand, $H_{\rm KE}$ 
and $H_{\rm SOC}$ 
are given as
\begin{subequations}
\begin{equation}
H_{\rm KE} =
\begin{pmatrix}
- \nabla^2 & 0 &0 \\
0 &- \nabla^2&0 \\
0 & 0 & - \nabla^2
\end{pmatrix},
\end{equation}{\label{KE3matrix}}
\begin{equation}
H_{\rm SOC} =-{\sqrt{2}}i \begin{pmatrix}
{\sqrt{2}\gamma_z}\partial_{z} & \gamma_x\partial_{x}
-i\gamma_y\partial_{y}&0 \\
\gamma_x\partial_{x}+i\gamma_y\partial_{y} & 0 & 
\gamma_x\partial_{x}-i\gamma_y\partial_{y}
\\
0 & \gamma_x\partial_{x}+i\gamma_y\partial_{y}  & 
-{\sqrt{2}\gamma_z}\partial_{z}
\end{pmatrix}{\label{SOC3matrix}}\\
\end{equation}
\end{subequations}
Since $H_{\rm SOC}$ can considered to be consisting of sum of two commuting 
Hamiltonians, i.e.,
\begin{eqnarray}
 H_{\rm SOC} &=& -{\sqrt{2}}i \left[\begin{pmatrix}
0 & \gamma_x\partial_{x}-i\gamma_y\partial_{y}&0 \\
\gamma_x\partial_{x}+i\gamma_y\partial_{y} & 0 & 
\gamma_x\partial_{x}-i\gamma_y\partial_{y}
\\0 & \gamma_x\partial_{x}+i\gamma_y\partial_{y}  & 0
\end{pmatrix} + \sqrt{2}\gamma_z\begin{pmatrix}
\partial_{z} &0&0 \\
0 & 0 & 0
\\
0 & 0  & -\partial_{z}
\end{pmatrix}\right],\nonumber\\
&=&H_{xy} + H_{z}.
\end{eqnarray}
With this division of $H_{\rm SOC}$ in 3D case, $H_{xy}$ becomes identical to 
$H_{\rm SOC}$ defined in Eq. ({\ref{SOC2matrix}}) for q2D case. 
The second Hamiltonian $H_z$ being diagonal can be combined with $H_{\rm KE}$. 
In other words, we can redefine $H_{\rm SOC}$ as 
simply $H_{xy}$ and $H_{\rm KE}$ as follows
 \begin{equation}
 H_{\rm KE} =
 \begin{pmatrix}
 - \nabla^2-{2}i\gamma_z\partial_{z} & 0 &0 \\
 0 &- \nabla^2&0 \\
 0 & 0 & - \nabla^2+{2}i\gamma_z\partial_{z}
 \end{pmatrix}.\label{KE3modmatrix}
 \end{equation}
The advantage of this redefining $H_{\rm SOC}$ making it 
identical to $H_{\rm SOC}$
in q2D case is that solution to equation (\ref{SOCpart}) is again given by
Eq. (\ref{sol_to_soc2d}) in Fourier space, whereas solution to 
Eq. (\ref{KEpart}) with $H_{\rm KE}$ defined in Eq. (\ref{KE3modmatrix}) is 
given by
 \begin{eqnarray}
 \hat{\phi}_{\pm 1}(k_x,k_y,k_z,t+\delta t) &=& 
\hat{\phi}_{\pm 1}(k_x,k_y,k_z,t)\exp[-i (k_x^2+k_y^2+k_z^2\pm 2\gamma_z k_z) 
\delta t],\nonumber\\
 \hat{\phi}_{0}(k_x,k_y,k_z,t+\delta t) &=& \hat{\phi}_{0}(k_x,k_y,k_z,t) 
\exp[-i (k_x^2+k_y^2+k_z^2) \delta t]\nonumber.
\end{eqnarray}

\subsection{Discretization Scheme}
In the current study, spin-1 BECs considered are either confined by external 
trapping potential or are self-localized by the interplay of the interactions
and spin-orbit coupling. This suggests that we can truncate our system from 
infinite space to some finite domain. In order to solve any equation 
computationally, we need to discretize our variables. We start by first 
truncating the spatial domain of the condensate along $\eta = x, y, z$ 
direction to $L_{\eta}$. Now, we choose 
$L_{\eta} =  N_{\eta}\times\Delta \eta$, where $\Delta \eta$ is the space-step
size chosen to discretize the spatial variable 
$\eta\in[-L_{\eta}/2,L_{\eta}/2]$ by setting 
$\eta_p = -L_{\eta}/2 + (p-1)\Delta \eta$ 
with $p = 1,2,\ldots,N_{\eta}+1$. The point $\eta_{N_{\eta}+1}$ is excluded 
from the set of the grid points due the periodicity of the wavefunction, 
$\phi (\eta_{N_{\eta}+1},t_q) = \phi(\eta_1,t_q)$. Similarly, time is 
discretized using $\Delta t$ as temporal step size. The discretization in the 
Fourier space which avoids the aliasing condition can be achieved by 
discretizing $k_{\eta}$ in $N_{\eta}$ equispaced $k_{\eta}$ points 
$\in \left[\frac{-N_{\eta} \pi}{L_{\eta}},
\ldots,\frac{(N_{\eta}-2) \pi}{L_{\eta}}\right]$ with a spacing of 
$2\pi/L_\eta$. The resultant discretized wavefunction 
$\phi_j(\eta_p,t_q)$ ($\hat{\phi}_j(k_\eta^p,t_q)$) in real (Fourier) space, 
where $p$ is the spatial (Fourier frequency) index and $q$ is the time index, 
make these amenable to be discrete Fourier transformed by FFTW software 
library (where ``in forward Fourier transform, positive frequencies are stored 
in the first half of the output and the negative frequencies are stored in 
backwards order in the second half of the output'') \cite{FFTW}, if 
$k_{\eta}^p$ are indexed as 
\begin{eqnarray}
k_{\eta}(i) &=& (i-1)\frac{2\pi}{L_{\eta}},
\quad i = 1,....\frac{N_{\eta}}{2}+1,\\
 k_{\eta}(i+1+\frac{N_{\eta}}{2}) &=
& -k_{\eta}(1-i+\frac{N_{\eta}}{2}),\quad i = 1,....\frac{N_{\eta}}{2}-1.
\end{eqnarray}
To summarize, the discrete analogues of the various continuous variables
are as follows:
 \begin{align}
       \eta &\equiv \eta_p,\quad
       k_\eta \equiv {k_\eta}^p,\quad
       t \equiv t_q,\\
       \phi_j(\eta,t) &\equiv \phi_j(\eta_p, t_q),\quad
       \hat{\phi_j}(k_\eta, t) \equiv \hat{\phi_j}({k_\eta}^p, t_q).
       \end{align}
The $N_\eta$ is chosen to be the multiple of $2$ to have the best performance 
from the FFTW subroutines \cite{FFTW}.

\subsection{Imaginary-time propagation }
 We use imaginary-time propagation, wherein $\delta t$ is replaced by 
$-i\delta t$, to compute the ground state of spin-1 BEC. This method neither 
preserves the norm nor the magnetization $\cal M$. To simultaneously fix 
the norm and magnetization, the component wavefunctions are redefined as
\begin{equation}
\phi_j(x^p,t_q) \equiv \sigma_j \phi_j(x^p,t_q),
\end{equation}
after each iteration in imaginary time where $\sigma_j$ are three projection 
parameters defined as 
\cite{bao2008computing}
 \begin{eqnarray}
 \sigma_0 &=& \frac{\sqrt{1-{\cal M}^2}}{[{N_0} + \sqrt{4(1-{\cal M}^2){N_1} 
{N_{-1}} +({\cal M} {N_0})^2}]^{1/2}},\\
  \sigma_1 &=& \sqrt{\frac{1+{\cal M}-( {\sigma_0} )^2 {N_0}}{2 {N_1}}},
\qquad {\sigma_{-1}} = \sqrt{\frac{1-{\cal M}
-( {\sigma_0} )^2 {N_0}}{2 {N_{-1}}}}.
 \end{eqnarray}
This simultaneous fixing of norm and $\cal M$ is not implemented in the 
presence of SO-coupling rather only the total norm is fixed. The reason being 
the existence of ground state solution with arbitrary magnetization is not 
guaranteed in this case. 

\section{Details about the programs} 
\label{section-IV}
In this section, we describe the set of three codes written in FORTRAN 90 
programming language. These three programs, namely  
\textbf{imretime\_spin1\_1D.f90}, \textbf{imretime\_spin1\_2D.f90}, and 
\textbf{imretime\_spin1\_3D.f90}, correspond to solving 1D 
Eqs. (\ref{cgpet1d-1})-(\ref{cgpet1d-3}), 2D 
Eqs. (\ref{cgpet2d-1})-(\ref{cgpet2d-3}) and 
3D Eqs. (\ref{cgpet-1})-(\ref{cgpet-3}), respectively, using the  
time-splitting spectral method described in the previous section. Each of 
these programs can solve the aforementioned equations with the user defined 
{\em option} of either imaginary-time or real-time propagation. 

The basic structure of the three codes is same; thus allowing us to describe 
the parameters, variables, modules, functions and subroutines using 1D code as
a prototypical example. 
\subsection{Modules}
First we provide the description of the four modules: 
{BASIC\_DATA}, CGPE\_DATA, SOC\_DATA, FFTW\_DATA.
\subsubsection*{BASIC\_DATA} The input parameters like the number of iterations 
(NITER), number of spatial-grid points (NX), spatial and temporal step sizes 
(DX and DT) are defined at the top of each program in this module. Besides 
these parameters, number of OpenMP/FFTW threads, constants like $\pi$ (PI), 
$i = \sqrt{-1}$ (CI), atomic mass unit (AMU), $\hbar$ (HBAR) and spatial 
domain $L_x$ (LX) are also defined in this module. 
\subsubsection*{CGPE\_DATA} The FORTRAN variables corresponding to $k_x$ (KX), 
$x$ (X), $V(x)$ (V), $a_{\rm osc}$ (AOSC), $\omega_x$ (OMEGAM), $c_0$ 
(C0), $c_2$ (C2), ${\cal M}$ (MAG), $\phi_j(x)$ (PHI), $\hat{\phi_j}(k_x)$ 
(PHIF) are declared in this module. The scattering lengths $a_0$ (A0), 
$a_2$ (A2); anisotropy parameters $\alpha_x$ (ALPHAX), 
$\alpha_y$ (ALPHAY), and $\alpha_z$ (ALPHAZ); mass $m$ (M) and 
total number of atoms $N$ (NATOMS) are defined in this module. 
In addition to this there are two user defined integer options: 
(a) SWITCH\_IM which has to be set equal to $1$ for imaginary-time propagation 
or 0 for real-time propagation and (b) OPTION\_FERRO\_POLAR 
which has to be set equal to 1, 2 or 3. OPTION\_FERRO\_POLAR = 1, 2 
correspond to suitable initial guess wavefunction for ferromagnetic and 
antiferromagnetic systems, respectively; whereas OPTION\_FERRO\_POLAR = 3
implies that the Gaussian initial guess wavefunctions would be used.

\subsubsection*{SOC\_DATA} The strength of spin-orbit coupling $\gamma_x$ (GAMMAX) is 
defined in this module. SWITCH\_SOC defined in this module has to be set equal
to $1$ if $\gamma_x \ne 0$ or equal to $0$ if $\gamma_x = 0$. {\bf The parameters 
and variables not listed in aforementioned three modules are not needed to be 
modified by the user.}

\subsubsection*{FFTW\_DATA} The variable types of the input and output arrays used in FFTW
subroutine to calculate discrete Fourier transform, requisite plans, and
thread initialization variable are declared in this module. The module uses
the FFTW3 module from the FFTW software library \cite{FFTW}, and is not
required to be modified by the user.

\subsection{Functions and subroutines} Now, we will describe the functions and subroutines which have been used in 
the programs.\\ 
{\em SIMPSON:} This function evaluates one-dimensional integral of form 
$\int f(x) dx$ using Simpson's 1/3 rule adapted for even number of grid points.
\\
\\
{\em DIFF}: This function evaluates $\partial f(x)/\partial x$ using nine 
point Richardson's extrapolation formula.
\\
\\
{\em INITIALIZE:} This subroutine initializes the initial guess wavefunctions 
PHI, space mesh X, trapping potential V, and Fourier frequencies KX. 
\\
\\
{\em NORMT}: The subroutine normalizes the total density to $1$.
\\
\\
{\em NORMC}: The subroutine calculates the norms of the individual components,
i.e. $\int |\phi_j(x)|^2 dx$.
\\
\\
{\em RAD}: The subroutine calculates the root mean square (rms) sizes of the three components.
\\
\\
{\em ENERGY}: The subroutine calculates the component chemical potentials 
$\mu_j$ (MU), $E$ (EN), and $\cal M$ (MAG).
\\
\\
{\em FFT} The subroutine calculates the discrete forward Fourier transform 
using freely available FFTW software library \cite{FFTW}. The subroutine uses 
the module FFTW3.
\\
\\
{\em BFT}: Similarly, the subroutine calculates the discrete backward Fourier 
transform using FFTW software library \cite{FFTW}. 
\\
\\
{\em KE}: The subroutine evaluates Eq. (\ref{sol_KEpartmom}) in Fourier space.
\\
\\
{\em SOC}: The subroutine implements Eq. (\ref{sol_to_soc}) with 
${\hat H}_{\rm SOC}$ given by Eq. (\ref{HSOC}).
\\
\\
{\em SE}: The subroutine implements Eq. (\ref{HSE_imp}) for $H_{\rm SE}$ 
consisting of spin-exchange terms.
\\
\\
{\em SP}: The subroutine implements Eq. (\ref{HSP_imp}) for $H_{\rm SP}$ 
consisting of spin-preserving terms.
\\
\\


  
 \subsection{2D and 3D programs}
As compared to 1D program which has NX grid points with spacing of DX, 
the 2D program requires NX $\times$ NY grid points with uniform spacing of DX 
and DY along $x$ and $y$ directions. This translates into spatial domain along
the two directions as LX = DX $\times$ NX,  LY = DY $\times$ NY. Similarly, 3D
program requires NX$\times$ NY$\times$ NZ grid points with corresponding space
steps of DX, DY and DZ. The spatial domain along three directions here is 
LX = DX $\times$ NX,  LY = DY $\times$ NY,  LZ = DZ $\times$ NZ. The additional
space variables Y and/or Z would also require corresponding Fourier frequencies
KY and/or KZ in 2D and 3D codes. The role of various subroutines is the direct
extension of the roles played by them in 1D code as per the discussion in 
sections \ref{q2D_section} and \ref{3D_section}.

 \subsection{Running the programs}
 One has to install FORTRAN compiler(s) and FFTW software library on the 
computer. If user is interested in finding the ground state of the spin-1 BEC,
the imaginary-time propagation has to be used. The dynamics on the other hand 
can be studied by real-time propagation using initial wave function which 
needs to be supplied by the user in the file 'initial\_sol.dat'. The 
compilation commands are listed at the top of each program file and also
in the `README.txt' file provided with the codes.

 \subsection{Description of Output files}
Data is written in four files during and after the execution of the 1D or 2D 
programs is complete. In the imaginary-time propagation, total norm, rms sizes
of the components, energy, absolute values of component wavefunctions at 
origin, and magnetization are written after every NSTP iterations, which is 
defined in the BASIC\_DATA module, in the file ``file1\_im.dat". 
In file ``file2\_im.dat", energy, chemical potentials, and rms sizes corresponding 
to each component are written after every STP iterations which is equivalent 
$0.1$ (dimensionless) time period. 
In the file ``tmp\_solution\_file.dat", which is updated after 
each NSTP iterations, component densities $\rho_j$ and corresponding phases are 
written at every space point. The final $\rho_j$ and corresponding phases are 
written in ``solution\_file\_im.dat". In real-time-propagation, the corresponding file 
names are `file1\_re.dat', `file2\_re.dat' and so on. There is another file, 
namely ``convergence.dat" which is written only in imaginary-time propagation.
In this file 
$\max |\phi_j(x_p,t_q) - \phi_j(x_p,t_q-\Delta t)|/(2 \Delta t)$ where
$-L_x/2\le x_p< L_x/2, i = -1,0,1$ and $t_q$ is the discrete imaginary-time 
is written after each iteration. This quantity serves as suitable convergence 
parameter, and the execution of the program is stopped if it falls below a 
user defined tolerance (TOL) defined in the CGPE\_DATA module. {\em For all
the results presented in this work a convergence tolerance of $10^{-6}$ has been met.}

In 3D code, besides the aforementioned four files, reduced densities in 
$x-y$ and $x-z$ planes and the corresponding phases are written in the files 
``tmp\_solution\_file\_xy.dat" and \\``tmp\_solution\_file\_xz.dat", 
respectively.

\subsection{Output samples from the codes}
Here we present the details of sample output files `file1\_im.dat' and/or `file1\_re.dat'
obtained from three codes. The contents of this file written in the 
successive filled lines are: (1) time stamp at the time of start; (2) number of OpenMP and FFTW 
threads used in the run; (3) values of SWITCH\_IM, OPTION\_FERROPOLAR,
SWITCH\_SOC, SO coupling strengths (GAMMAX, GAMMAY, GAMMAZ), and tolerance (TOL) 
used; (4) values of anisotropy parameters (ALPHAX, ALPHAY, ALPHAZ) chosen; 
(5) number of space grid points (NX, NY, NZ); (6) values of NITER and NSTP;
(7) value of space step(s) (DX, DY, DZ), (7) time step DT, space domain (LX, LY, LZ), 
and magnetization (MAG); (8) frequency used in scaling (OMEGAM), corresponding
oscillator length (AOSC), and values of interaction parameters (C0, C2).
Then total norm, rms sizes of the component wavefunctions, energy, absolute
values of component wavefunctions at the origin, and magnetization are written
for initial solution, for the transient solution obtained after NSTP time 
iterations and for the converged solution (this third entry in real-time code will simply
correspond to the solution after NITER iterations). The time stamp at the end
of the run and execution time are the last two entries in this file.
The varied nature of the contents of this file can be used to ascertain the 
success of the run of the code by verifying the input parameters selected and 
various output parameters. The sample output files obtained with the test runs
of imretime\_spin1\_1D.f90, imretime\_spin1\_2D.f90, and imretime\_spin1\_3D.f90
are presented in the Electronic Appendix.
In all the test runs, harmonic trapping potential as per
the trapping potential corresponding to anisotropy parameters listed
in these files have been used. For these test runs, the codes were compiled with 
Intel's FORTRAN compiler and the jobs were run on a server with  
two Intel\textsuperscript{\textregistered} Xeon\textsuperscript{\textregistered} 
Platinum 8180 CPU @ 2.50GHz. The samples of all the data files, both input
and output, corresponding to the current set of parameters in the codes are available
on Mendeley data \cite{FORTRESS-DATA}.

\section{OpenMP Parallelization}\label{section-V}

\begin{figure}[H]
 \includegraphics[trim = 0mm 0mm 0mm 0mm,clip, width=1.0\linewidth,clip]
 {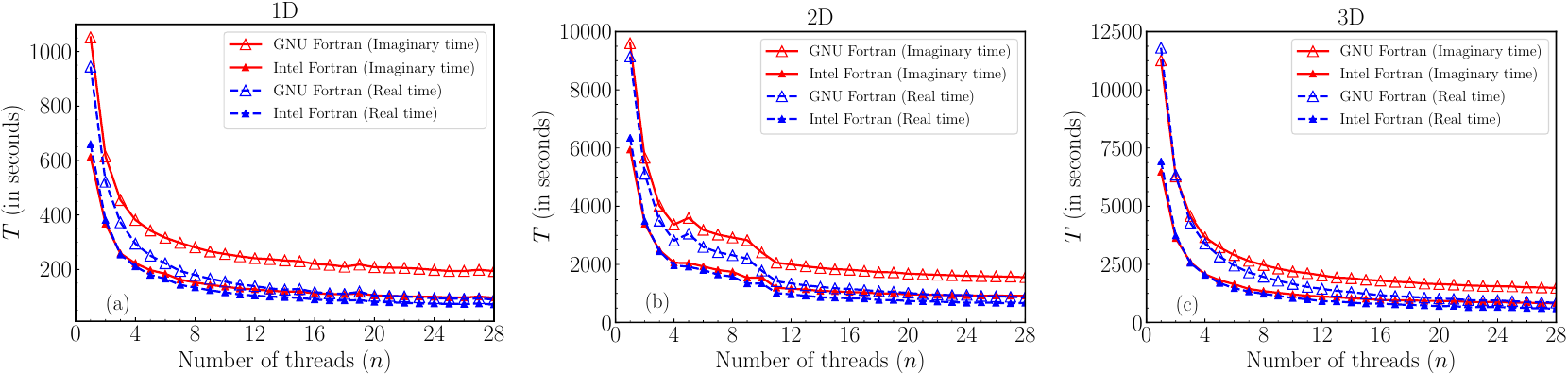}
 \caption{ (Color online) (a) Execution times for 1000 iterations (in seconds) as 
function of number of threads for 1D code compiled with GNU Fortran 5.4.0 and 
Intel Fortran 19.1.0.166 compilers for imaginary and real-time propagation. 
(b) and (c) are the same for 2D and 3D codes.}
 \label{execution_times}
 \end{figure}

We have tested the efficiency of OpenMP parallelization of the three codes
for both imaginary and real-time propagations. The tests were done on
a 28-core Intel\textsuperscript{\textregistered} Xeon\textsuperscript{\textregistered} 
Platinum 8180 CPU @ 2.50GHz processor.
The parallelization tests were performed with NX = $10^6$ for 1D code,
NX = NY = $3000$ for 2D code, and NX = NY = $256$, NZ = $128$ for 3D code. 
The execution time was measured for 1000 iterations starting from the call 
to INITIALIZE subroutine and did not include the time spent in reading/writing
and opening/closing the data files. The execution times $T(n)$ for the three codes compiled with both GNU 
Fortran 5.4.0 and Intel Fortran 19.1.0.166 compilers are shown as a function
of number of threads $n$ in Fig. \ref{execution_times}. It is evident from 
Fig. \ref{execution_times} that the codes compiled with Intel Fortran compiler
are faster than those compiled with GNU Fortran compiler for both
the imaginary time and real-time propagations; nonetheless the difference
in the execution times for codes compiled with these two compilers is
less for real-time propagation. The execution times in all the cases
shown in Fig. \ref{execution_times} first decrease very sharply with the
increase in the number of threads and then tends to saturate with 
increasing number of threads.

\begin{figure}[H]
\includegraphics[trim = 0mm 0mm 0mm 0mm,clip, width=1.00\linewidth,clip]{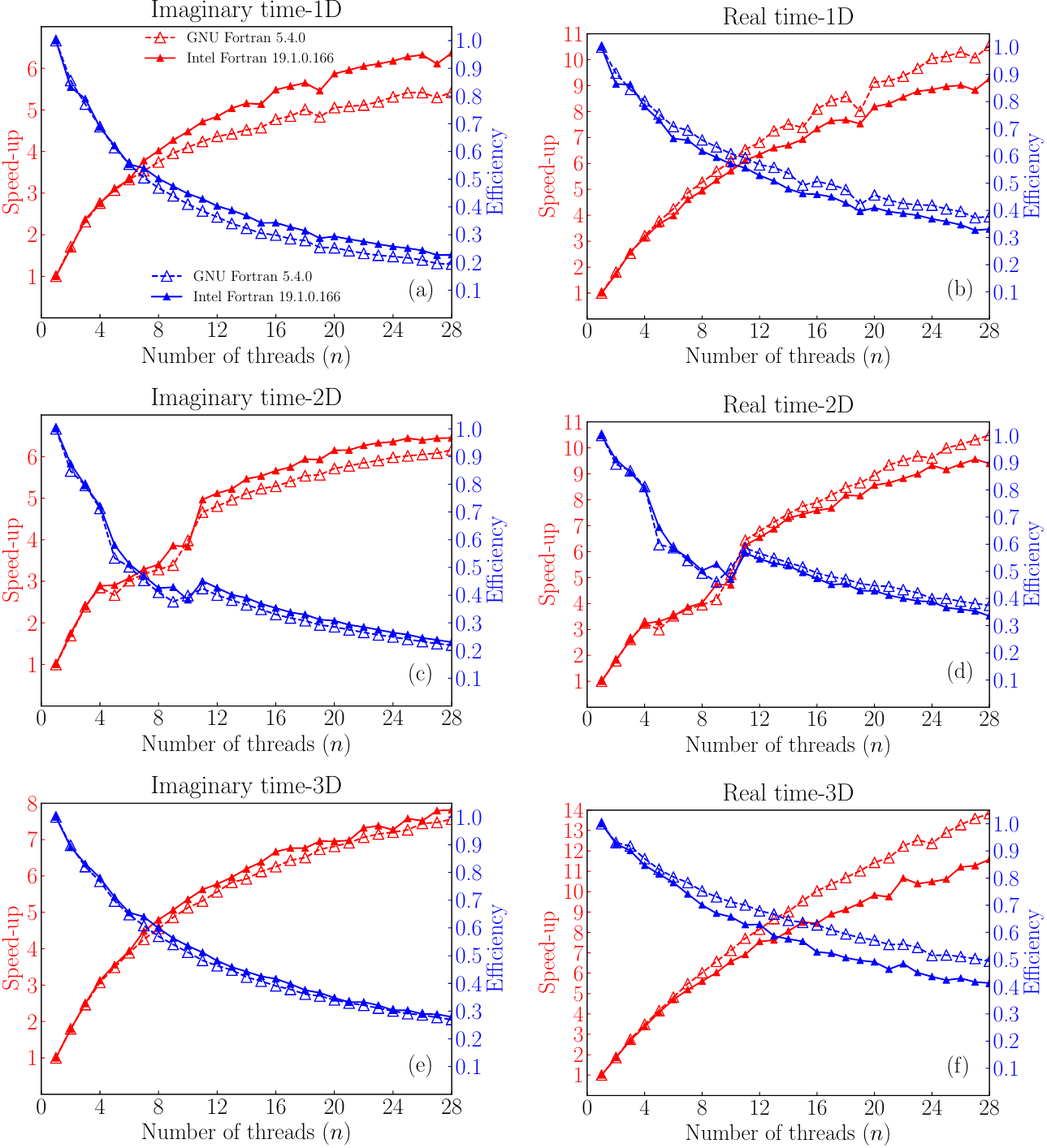}
\caption{(Color online) Speedup and efficiency as a function of number of threads $n$ 
are shown for imaginary time (left column) and real-time propagation (right column). 
Top, middle and bottom row figures show the results for 1D, 2D and 3D codes, respectively.
}
\label{speedup_eff}
\end{figure}

To quantify the performance gain with OpenMP parallelization, we have calculated the 
speedup and efficiency for all these codes compiled with the aforementioned two compilers. 
Here speedup is defined as the ratio of execution time with 1 thread to the execution time 
with $n$ threads, i.e $T(1)/T(n)$, whereas the efficiency is defined as the
ratio $T(1)/[nT(n)]$. For all the codes, speedup and efficiency as a function of number 
of threads are much better for real-time propagation as compared to imaginary time propagation.
The real-time speedup achieved with 28 threads was more than 9 for both 1D and 2D codes,
and more than 11 for 3D code using both the compilers; whereas the corresponding imaginary
time speedup values are more than 5 for 1D, more than 6 for 2D and more than 7 for 3D with
both the compilers as is shown in Fig. \ref{speedup_eff}. The best performing real-time
3D has more than $40\%$ efficiency with $28$ threads. The better performance of real-time variants
is due to fact that the imaginary propagation has to fix the norm and
also has to check the convergence criterion during each iteration. 
Real-time propagation corresponds to the unitary evolution of a converged solution,
and hence does not need to fix the norm or check the convergence. All the results presented
in this section were performed for non-zero value of SO-coupling strength.

\section{Numerical Results}\label{section-VI}
In this section, we present the results for energy, chemical potentials, and 
densities of the ground states in q1D, q2D and 3D spin-1 condensates using 
the imaginary time propagation method with the emphasis on the comparison with the 
previously published results in the literature \cite{bao2008computing,bao2013efficient,
gautam2018three,gautam2017vortex,gautam2015mobile}. 
We report the results in the presence as well as absence of SO coupling. 
To check the accuracy of numerical method employed by us, we compare our 
results in the absence of SO coupling with those in 
Ref. \cite{bao2008computing,bao2013efficient}. In the presence of SO coupling,
we compare our results in q1D, q2D and 3D spin-1 BECs with those in Refs.
  \cite{gautam2018three}, \cite{gautam2017vortex} and \cite{gautam2015mobile}
respectively. It needs to emphasized that the method used in 
Ref. \cite{bao2013efficient} is not applicable to SO-coupled spin-1 BECs.

\subsection{Results for q1D spin-1 BECs}
\subsubsection{Without SO coupling, $\gamma_x = 0$}
We choose our computational domain $L = [-16,16]$ having spatial step size as 
$\Delta x = 1/64$ for q1D condensates. We first consider (a) ferromagnetic 
spin-1 BEC of $^{87}$Rb confined in a cigar-shaped trapping potential having 
interaction parameters in dimensionless units as $c_0 = 0.0885N$ and 
$c_2 = -0.00041N$ \cite{bao2008computing} and (b) antiferromagnetic spin-1 
condensate of $^{23}$Na confined in a cigar-shaped trapping potential having 
interaction parameters in dimensionless units as $c_0 = 0.0241N$ and 
$c_2 = 0.00075N$  \cite{bao2008computing,bao2013efficient} for our 
computations in one dimensional case. We consider $N = 10^4$ as the total 
number of atoms in each of these two cases. The comparison of ground state 
energies obtained in present work with those in Refs. 
\cite{bao2008computing,bao2013efficient} is excellent as is shown in table 
\ref{table1} for  $\Delta x \le 1/64$  and $\Delta t \approx 0.1{\Delta x}^2$
\begin{table}[!h]
\caption{Ground state energies for $^{87}$Rb and $^{23}$Na q1D BECs obtained 
in the present work with $\Delta x \leq 0.015625$ and $\Delta t \approx 0.1 
{\Delta x}^2$ along with the same from Ref \cite{bao2008computing} for the 
various values of magnetization $\cal M$.}
\begin{center}
\begin{tabular}{| c | c | c | c | c |}
\hline
\multicolumn{1}{ |c| }{}&\multicolumn{2}{|c|}{$^{87}$Rb}&
\multicolumn{2}{ |c| }{$^{23}$Na}\\
\hline
\multicolumn{1}{ |c| }{${\cal M}$}& 	
\multicolumn{1}{ |c| }{$E$ in Ref.\cite{bao2008computing}} & 		
\multicolumn{1}{ |c| }{$E$ (present work)}& 		
\multicolumn{1}{ |c| }{$E$ in Ref. \cite{bao2008computing,bao2013efficient}} &
\multicolumn{1}{ |c| } {$E$ (present work)} \\
\hline
0   &36.1365 &  36.1365 & 15.2485  &15.2485\\ \hline
0.1 &36.1365 &  36.1365 & 15.2513  &15.2513\\ \hline
0.2 &36.1365 &  36.1365 & 15.2599  &15.2599\\ \hline
0.3 &36.1365 &  36.1365 & 15.2743  &15.2743\\ \hline
0.4 &36.1365 &  36.1365 & 15.2945  &15.2945\\ \hline
0.5 &36.1365 &  36.1365 & 15.3209  &15.3209\\ \hline
0.6 &36.1365 &  36.1365 & 15.3537  &15.3537\\ \hline
0.7 &36.1365 &  36.1365 & 15.3933  &15.3933\\ \hline
0.8 &36.1365 &  36.1365 & 15.4405  &15.4405\\ \hline
0.9 &36.1365 &  36.1365 & 15.4962  &15.4962\\ \hline
\end{tabular}
\label{table1}
\end{center}
\end{table}

For q1D $^{87}$Rb, we also consider an alternative set of interaction 
parameters of $c_0 = 0.08716N$, $c_2 = -0.001748N$ and $N = 10000$ for our 
computations \cite{bao2013efficient}. In this case again, the ground state 
energy obtained in the present work is in excellent agreement with the value 
reported in Ref. \cite{bao2013efficient} as is shown in table \ref{table2} 
for the same interaction parameters set.
\begin{table}[H]
\caption{Comparison of the ground state energy of q1D $^{87}$Rb condensate 
reported in Ref. {\cite{bao2013efficient}} with the value obtained in the 
present work with $\Delta x \leq 0.0156255$ and 
$\Delta t \approx 0.1\Delta x^2$}
\begin{center}
\begin{tabular}{| c | c | c |}
\hline
\multicolumn{1}{ |c| }{}&\multicolumn{2}{ |c| }{$^{87}$Rb} \\
\hline
\multicolumn{1}{ |c| }{${\cal M}$} & 	
\multicolumn{1}{ |c| }{$E$ in Ref. \cite{bao2013efficient}} & 		
\multicolumn{1}{ |c| }{$E$ (present work)} \\
\hline
0-0.9 & 35.4007  &35.4007[7] \\ \hline
\end{tabular}
\label{table2}
\end{center}
\end{table}

The chemical potential values obtained in present work are also in very good
agreement with those reported in Ref. \cite{bao2008computing} as is shown in 
table \ref{table3}. 
\begin{table}[H]
\caption{Comparison of the chemical potential values for $^{87}$Rb and 
$^{23}$Na condensate reported in Ref. \cite{bao2008computing} with the values 
obtained in the present work with $\Delta x = 0.0025$, $\Delta t = 0.0000095$.
For $^{23}$Na, $\mu = (\mu_{+1} + \mu_{-1})/2$, whereas for $^{87}$Rb 
$\mu = \mu_0 = \mu_{\pm 1}$.} 
\begin{center}
\begin{tabular}{| c | c | c | c | c |}
\hline
\multicolumn{1}{ |c| }{}&
\multicolumn{2}{|c|}{$^{87}$Rb}&
\multicolumn{2}{ |c| }{$^{23}$Na}\\
\hline
\multicolumn{1}{ |c| }{${\cal M}$} & 	
\multicolumn{1}{ |c| }{$\mu$ in Ref. \cite{bao2008computing}} &
\multicolumn{1}{ |c| }{$\mu$ (present work)}& 		
\multicolumn{1}{ |c| } {$\mu$ in \cite{bao2008computing}} &
\multicolumn{1}{ |c| } {$\mu$ (present work)} \\
\hline
0   &60.2139 &  60.2136 & 25.3857  &25.3857\\ \hline
0.1 &60.2139 &  60.2136 & 25.3847  &25.3838\\ \hline
0.2 &60.2139 &  60.2136 & 25.3815  &25.3804\\ \hline
0.3 &60.2139 &  60.2136 & 25.3762  &25.3749\\ \hline
0.4 &60.2139 &  60.2137 & 25.3682  &25.3668\\ \hline
0.5 &60.2139 &  60.2137 & 25.3572  &25.3557\\ \hline
0.6 &60.2139 &  60.2137 & 25.3423  &25.3406\\ \hline
0.7 &60.2139 &  60.2138 & 25.3220  &25.3203\\ \hline
0.8 &60.2139 &  60.2138 & 25.2939  &25.2921\\ \hline
0.9 &60.2139 &  60.2139 & 25.2527  &25.2509\\ \hline
\end{tabular}
\label{table3}
\end{center}
\end{table}

\begin{figure}[H]
\includegraphics[trim = 5mm 0mm 4.5mm 0mm,clip, width=0.56\linewidth,clip]
{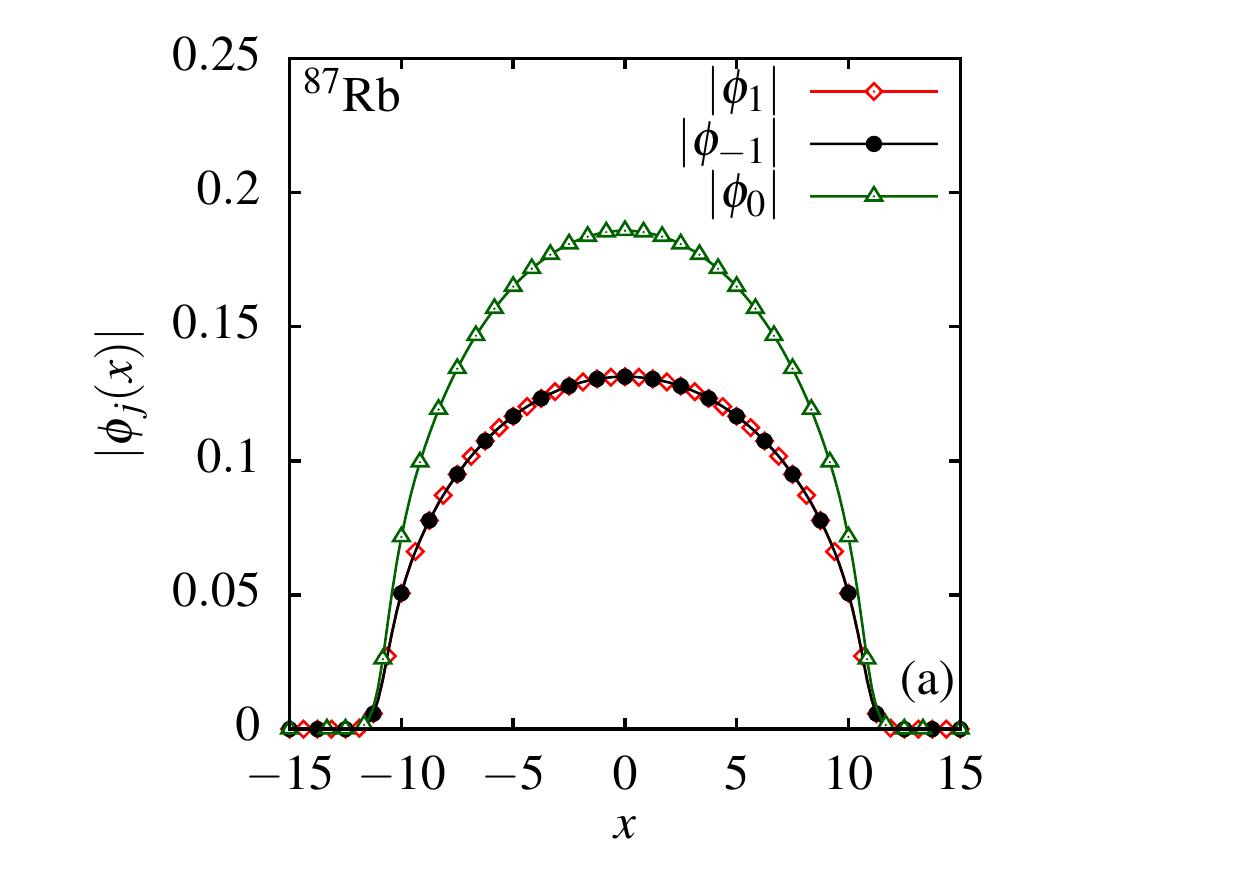}\hfill
 \hspace{0.1cm}
 \includegraphics[trim = 5mm 0mm 4.5mm 0mm,clip, width=0.56\linewidth,clip]
{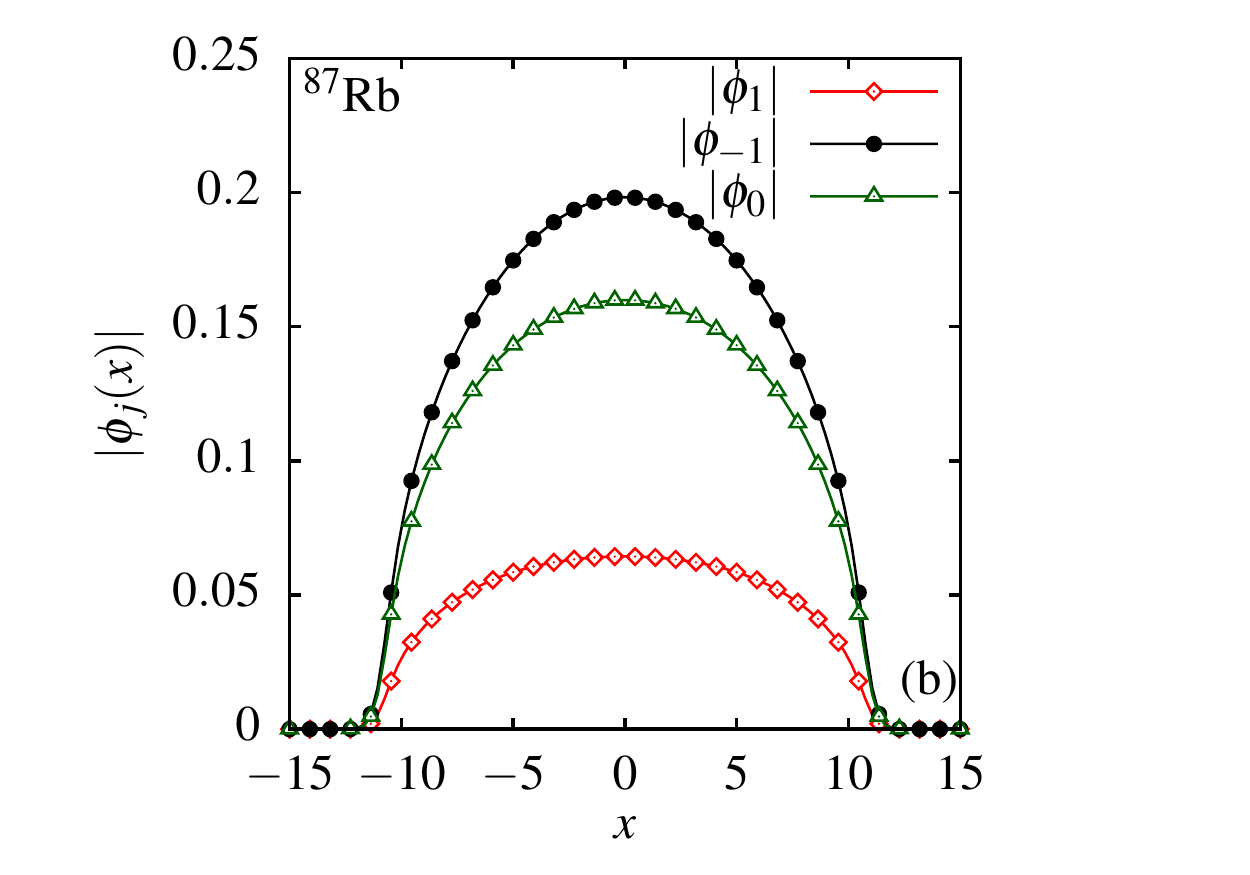}\hfill
 \includegraphics[trim = 5mm 0mm 4.5mm 0mm,clip, width=0.56\linewidth,clip]
{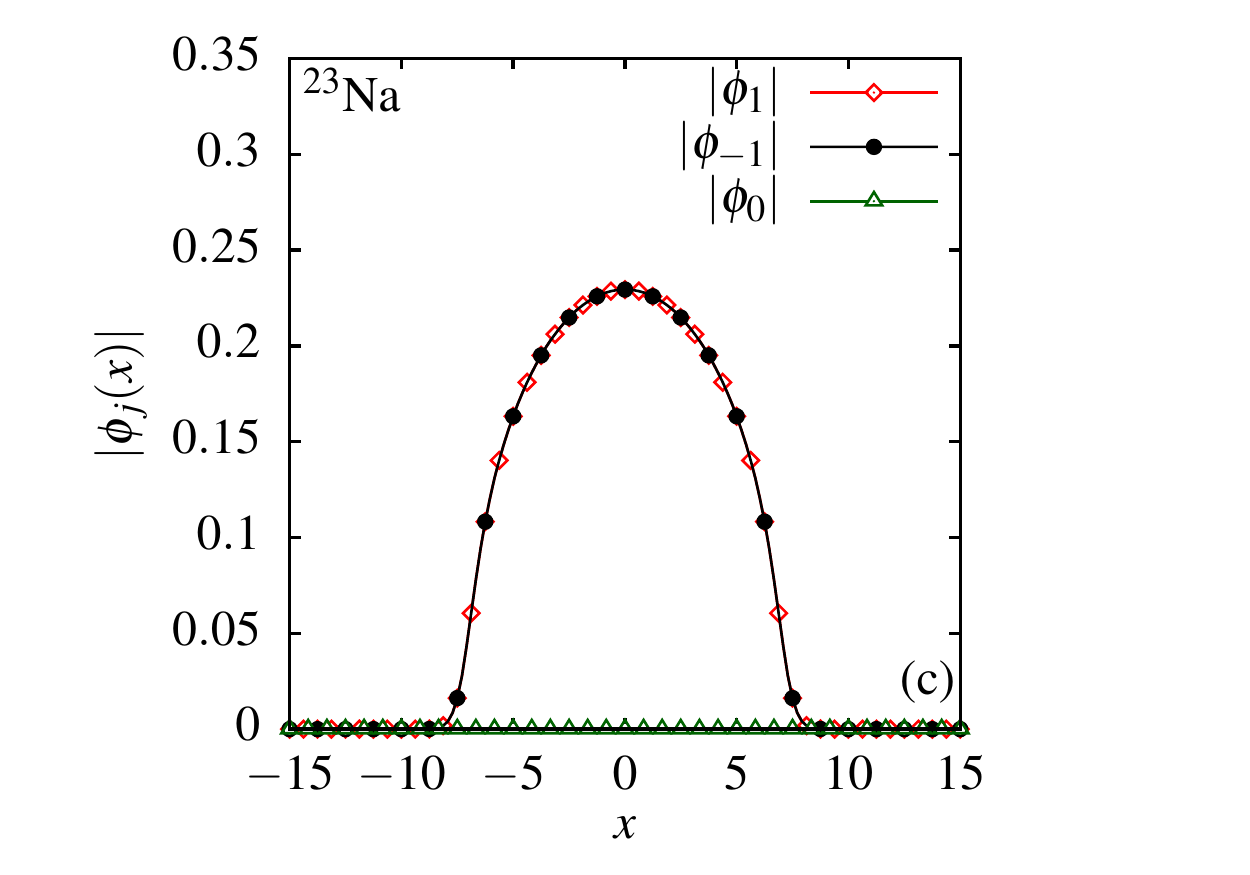}\hfill
 \hspace{0.1cm}
 \includegraphics[trim = 5mm 0mm 4.5mm 0mm,clip, width=0.56\linewidth,clip]
{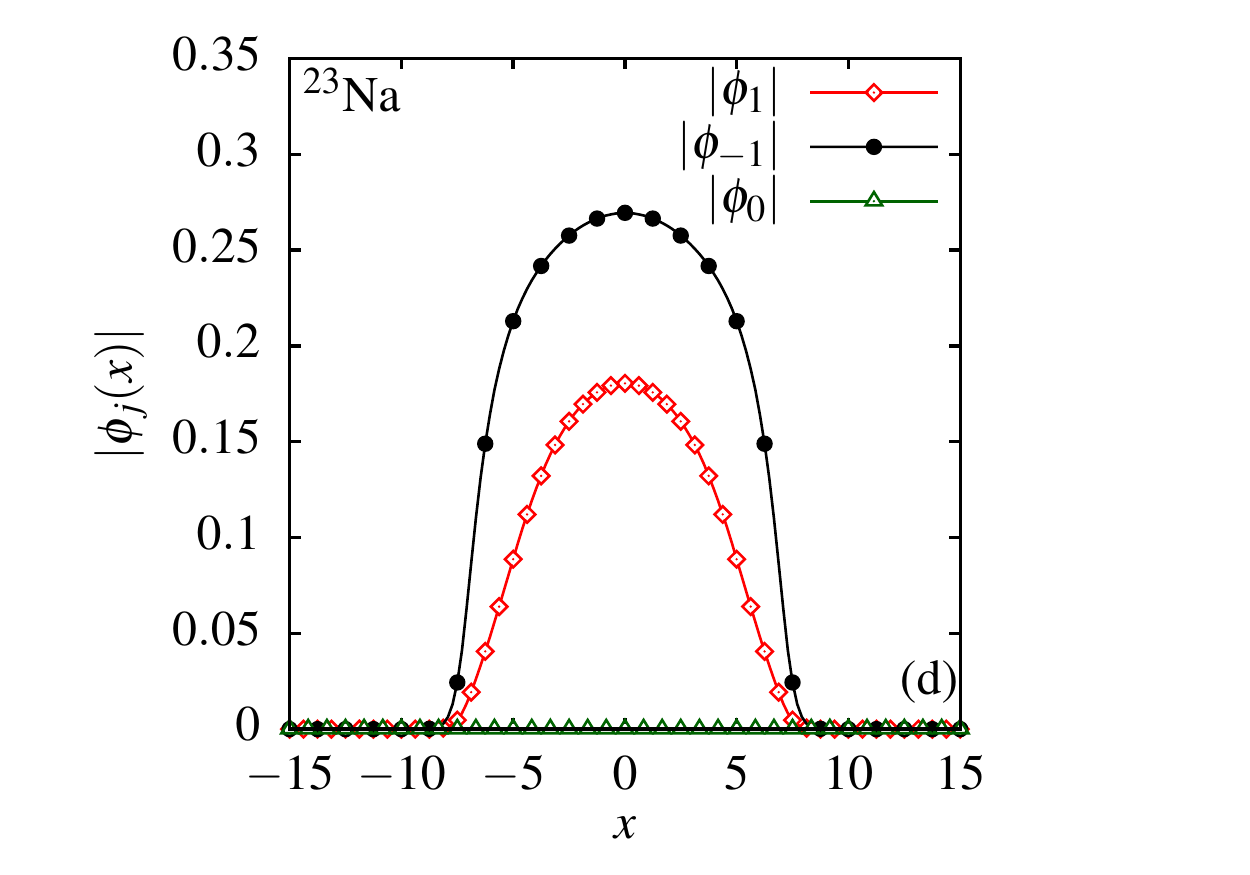}\hfill
 \caption{ (Color online) Absolute values of component wavefunctions 
$|\phi_j(x)|$ in the ground state of $^{87}$Rb for (a) ${\cal M} = 0$, 
(b)  ${\cal M} = 0.5$. (c) and (d) are the same for $^{23}$Na with 
${\cal M} = 0$  and ${\cal M} = 0.5$, respectively. 
These are in agreement with \cite{bao2008computing,bao2013efficient}}.
\label{fig1}
\end{figure}

The ground state wavefunctions are also in excellent agreement with Ref. 
\cite{bao2008computing}. The absolute values of ground state wavefunctions 
for $^{87}$Rb and $^{23}$Na with ${\cal M} = 0$ and $0.5$ are shown in 
Fig. {\ref{fig1}}.

\subsubsection{With SO coupling, $\gamma_x\ne 0$}
\begin{figure}[H]

        \includegraphics[trim = 5mm 0mm 4.5mm 0mm,clip, width=0.56\linewidth,clip]{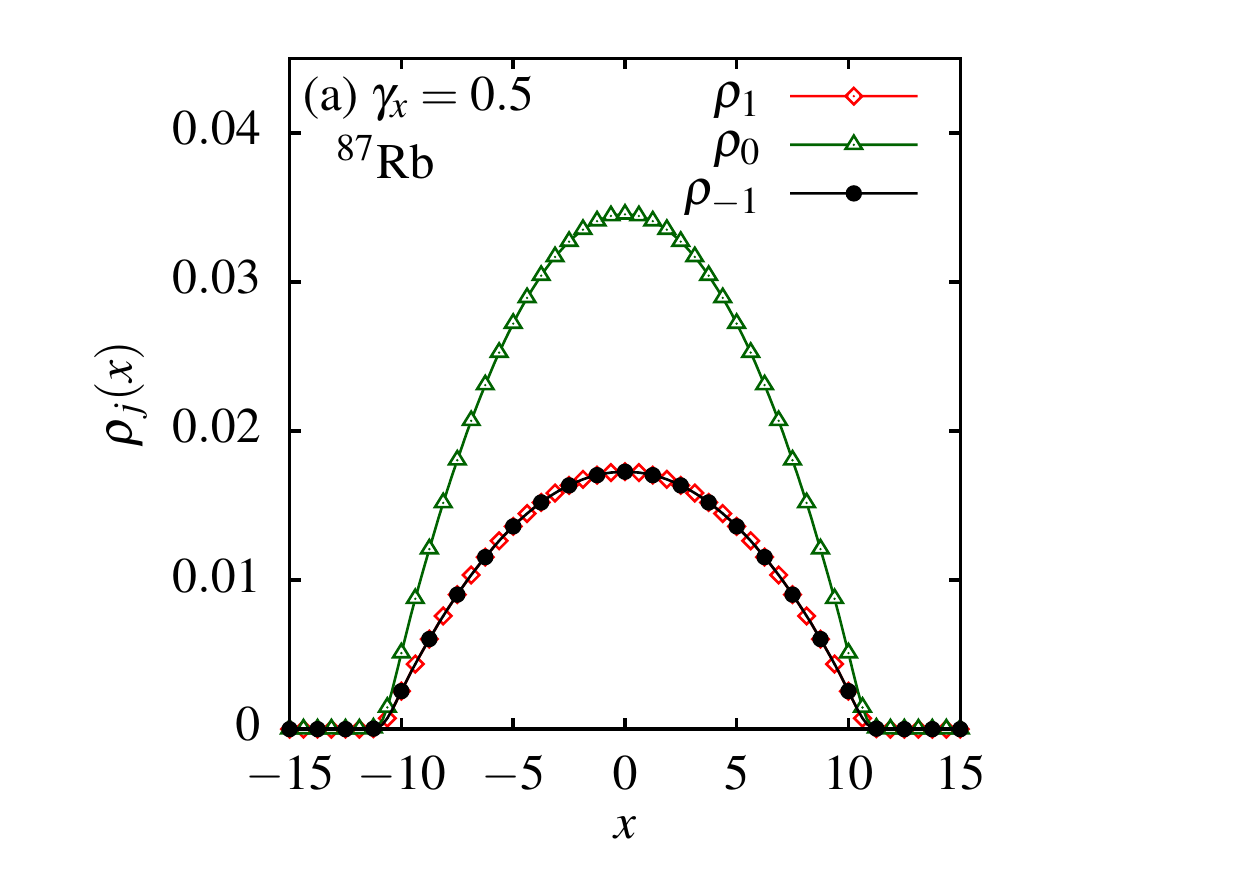}
        \includegraphics[trim = 5mm 0mm 4.5mm 0mm,clip, width=0.56\linewidth,clip]{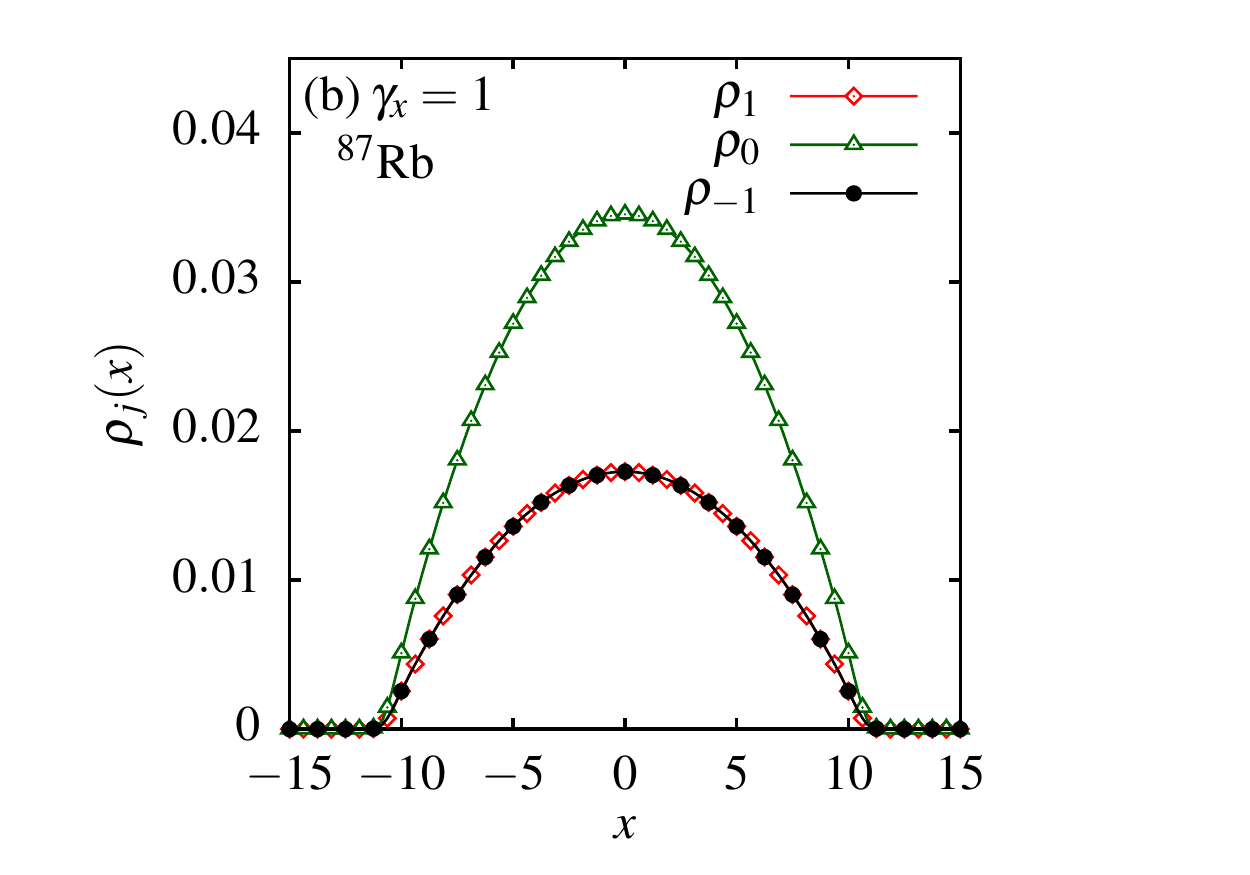}
        \hspace{0.1cm}
        \includegraphics[trim = 5mm 0mm 4.5mm 0mm,clip, width=0.56\linewidth,clip]{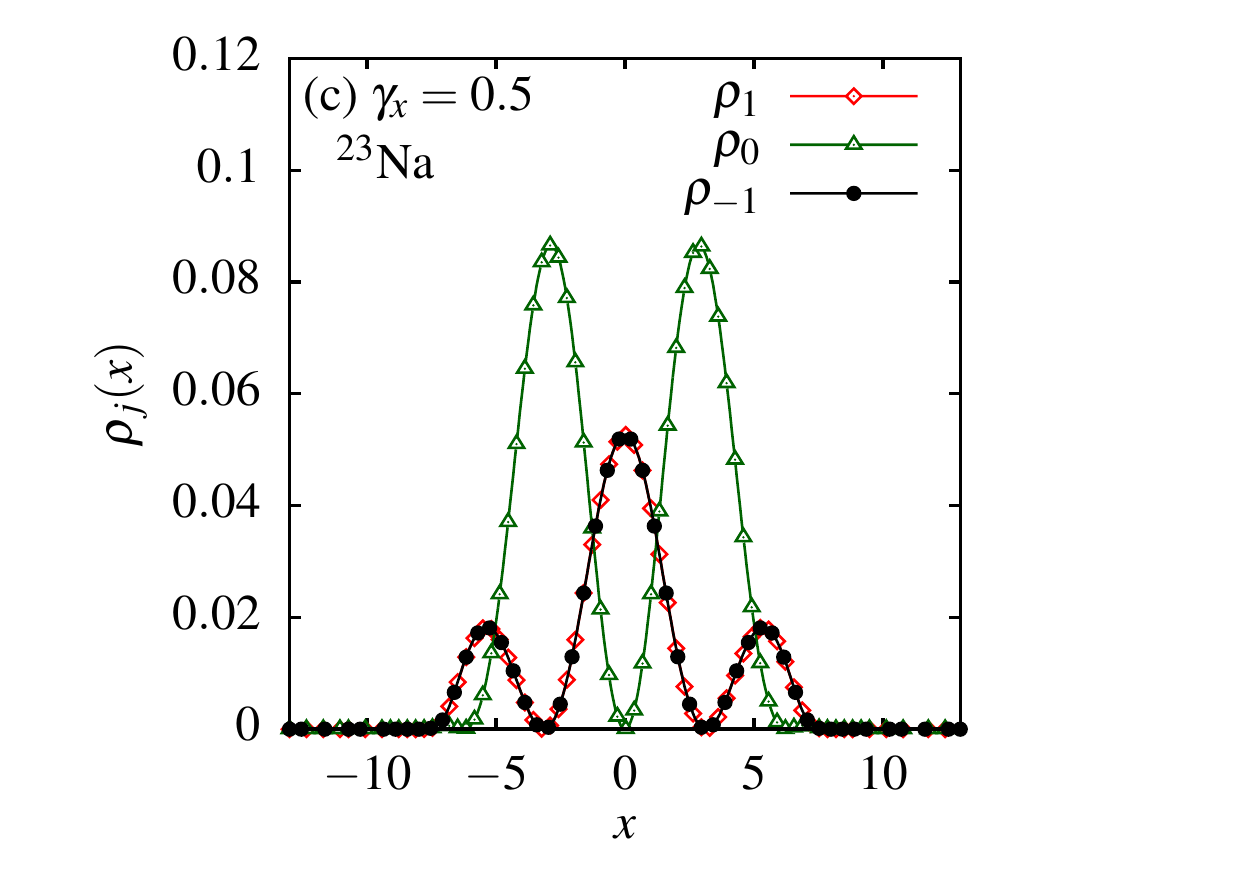}
        \includegraphics[trim = 5mm 0mm 4.5mm 0mm,clip, width=0.56\linewidth,clip]{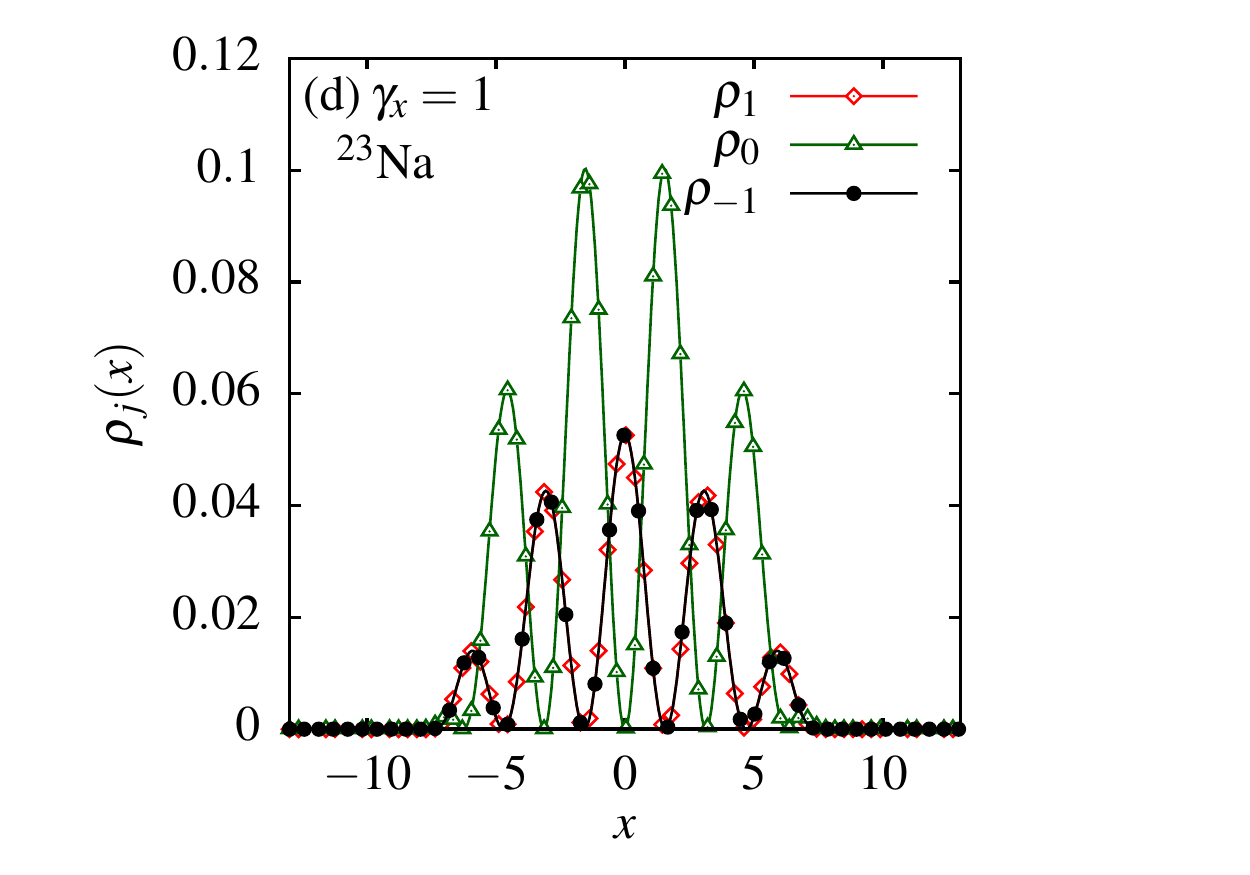}
        \hspace{0.1cm}
        \caption{ (Color online) Ground state density of SO-coupled $^{87}$Rb 
for (a) $\gamma_x = 0.5$, (b) $\gamma_x = 1$. (c) and (d) are the same for 
$^{23}$Na with $\gamma_x = 0.5$ and $\gamma_x = 1.0$, respectively. ${\cal M} = 0$
in all the cases.}
        \label{fig2}
\end{figure}

In the presence of SO coupling with harmonic trapping potential, 
for $^{87}$Rb and $^{23}$Na, we again consider $(c_0,c_2)$ equal to 
$(0.08716 N, -0.001748N)$ and $(0.0241N,0.00075N)$, respectively, 
where $N = 10000$. The ground state energy values in these cases are given in 
table \ref{table4} for multiple values of $\gamma_x$.
\begin{table}[!h]
\caption{Ground state energies of $^{87}$Rb and $^{23}$Na condensates in the 
presence of harmonic trap and spin-orbit coupling with 
$\Delta x = 0.015625$ , $\Delta t \approx  0.1(\Delta x)^2$. The $(c_0,c_2)$
values are $(0.08716 N,-0.001748N)$ and  $(0.0241N,0.00075N)$ with $N =10000$ 
for $^{87}$Rb and $^{23}$Na, respectively.}
\begin{center}
\begin{tabular}{| c | c | c |}
\hline
\multicolumn{1}{ |c| }{}&\multicolumn{1}{|c|}{$^{87}$Rb}&
\multicolumn{1}{ |c| }{$^{23}$Na}\\
\hline
\multicolumn{1}{ |c| }{$\gamma_x$}&\multicolumn{1}{|c|}{Energy}&
\multicolumn{1}{ |c| }{Energy}\\
\hline
0    & 35.4007 &  15.2485  \\ \hline
0.1  & 35.3958 &  15.2435 \\ \hline
0.2  & 35.3808 &  15.2285  \\ \hline
0.3  & 35.3558 &  15.2035 \\ \hline
0.4  & 35.3208 &  15.1685 \\ \hline
0.5  & 35.2758 &  15.1235 \\ \hline
0.6  & 35.2208 &  15.0685 \\ \hline
0.7  & 35.1558 &  15.0035 \\ \hline
0.8  & 35.0808 &  14.9285 \\ \hline
0.9  & 34.9958 &  14.8435  \\ \hline
1   & 34.9008 &  14.7485 \\ \hline
\end{tabular}
\label{table4}
\end{center}
\end{table}
The component densities for the two systems with $\gamma_x = 0.5$ 
and $1$ are shown in Fig. \ref{fig2}.

\begin{figure}[H]
\includegraphics[trim = 5mm 0mm 4.5mm 0mm,clip, width=0.56\linewidth,clip]
{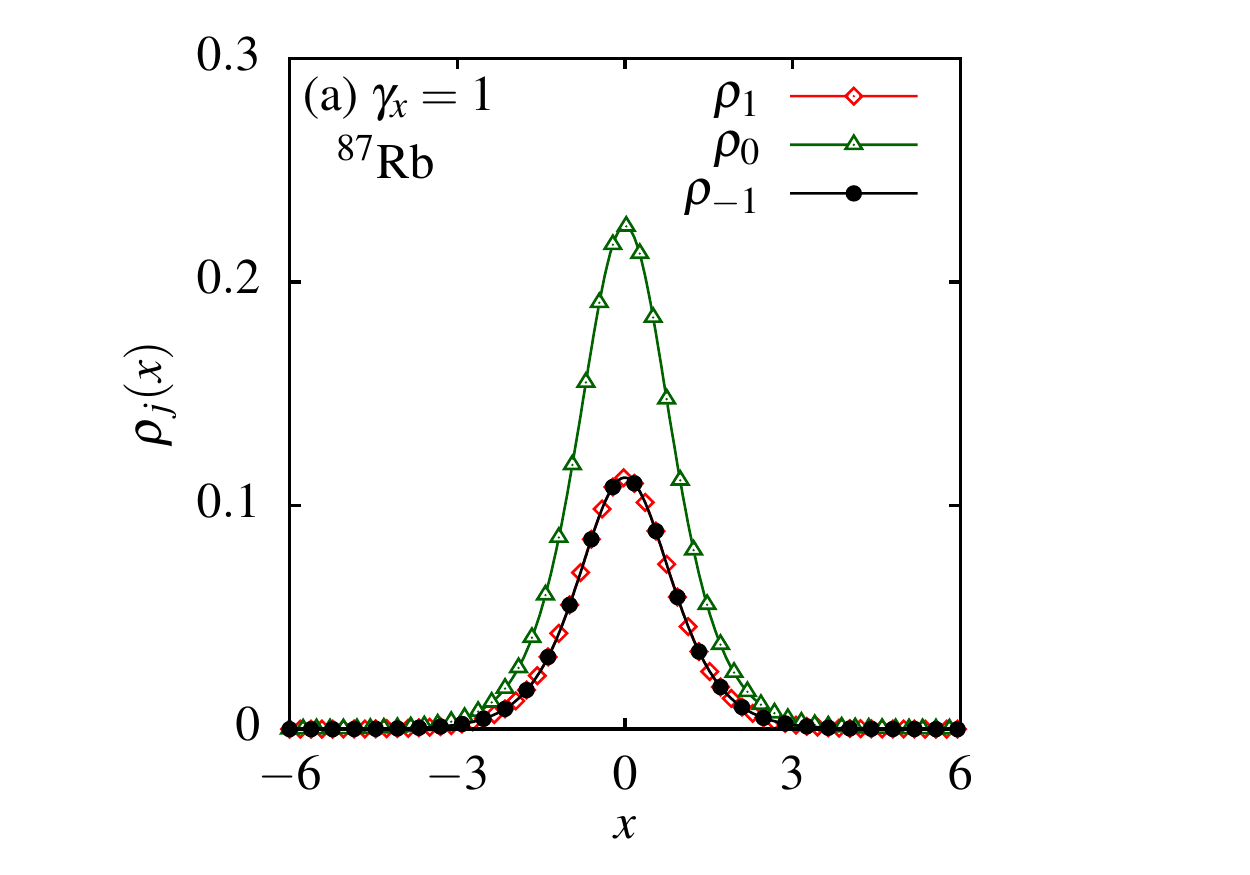}
\includegraphics[trim = 5mm 0mm 4.5mm 0mm,clip, width=0.56\linewidth,clip]
{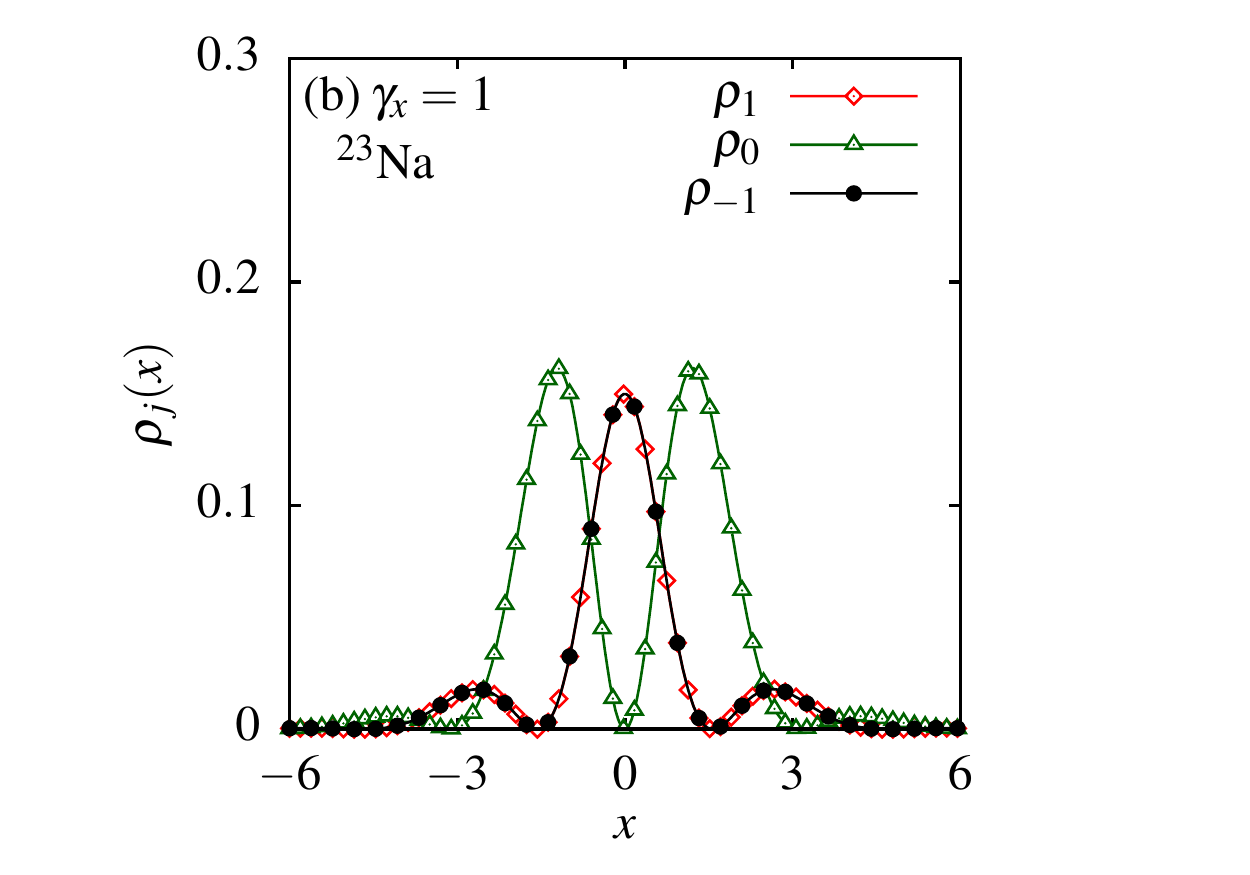}
\hspace{0.1cm}
\caption{ (Color online) (a) Ground state density of SO-coupled spin-1 BEC  
with $c_0 = -1.5, c_2 = -0.3$ in the absence of trap and $\gamma_x = 1$. 
(b) The same for $c_0 = -1.2, c_2 = 0.3$. These results are in agreement with 
Ref.  \cite{gautam2015mobile} and correspond to ${\cal M} = 0$.}
\label{fig3}
\end{figure}

Next, we consider ferromagnetic and antiferromagnetic systems with $(c_0,c_2)$
equal to $(-1.5,-0.3)$ and $(-1.2,0.3)$, respectively in the absence of any 
trapping. The ground state energies of the self-trapped solutions obtained in 
these cases are shown in table \ref{table5}. 
\begin{table}[!h]
\caption{Ground state energies for self-trapped ferromagnetic and 
antiferromagnetic condensates in the presence of spin-orbit coupling obtained 
with $\Delta x = 0.015625$, $\Delta t \approx  0.1 (\Delta x)^2$.} 
\begin{center}
\begin{tabular}{| c | c | c |}
\hline
\multicolumn{1}{ |c| }{}&
\multicolumn{1}{|c|}{$(c_0,c_2) = (-1.5,-0.3)$}&
\multicolumn{1}{ |c| }{$(c_0,c_2) =(-1.2,0.3)$}\\
\hline
\multicolumn{1}{ |c| }{$\gamma_x$}&\multicolumn{1}{|c|}{Energy}&
\multicolumn{1}{ |c| }{Energy}\\
\hline
0    & -0.1350 & -0.0600  \\ \hline
0.1  & -0.1400 & -0.0650  \\ \hline
0.2  & -0.1550 & -0.0800  \\ \hline
0.3  & -0.1800 & -0.1050  \\ \hline
0.4  & -0.2150 & -0.1400  \\ \hline
0.5  & -0.2600 & -0.1850  \\ \hline
0.6  & -0.3150 & -0.2400  \\ \hline
0.7  & -0.3800 & -0.3050  \\ \hline
0.8  & -0.4550 & -0.3800  \\ \hline
0.9  & -0.5400 & -0.4650  \\ \hline
1    & -0.6350 & -0.5600  \\ \hline
\end{tabular}
\label{table5}
\end{center}
\end{table}
The self-trapped nature of the solutions is evident from the ground state 
densities shown in Fig. \ref{fig3} for $\gamma_x = 1$.

\subsection{Real-time check}
\begin{figure}[H]
	\includegraphics[trim = 0mm 0mm 1cm 0mm,clip, width=0.55\linewidth,clip]
	{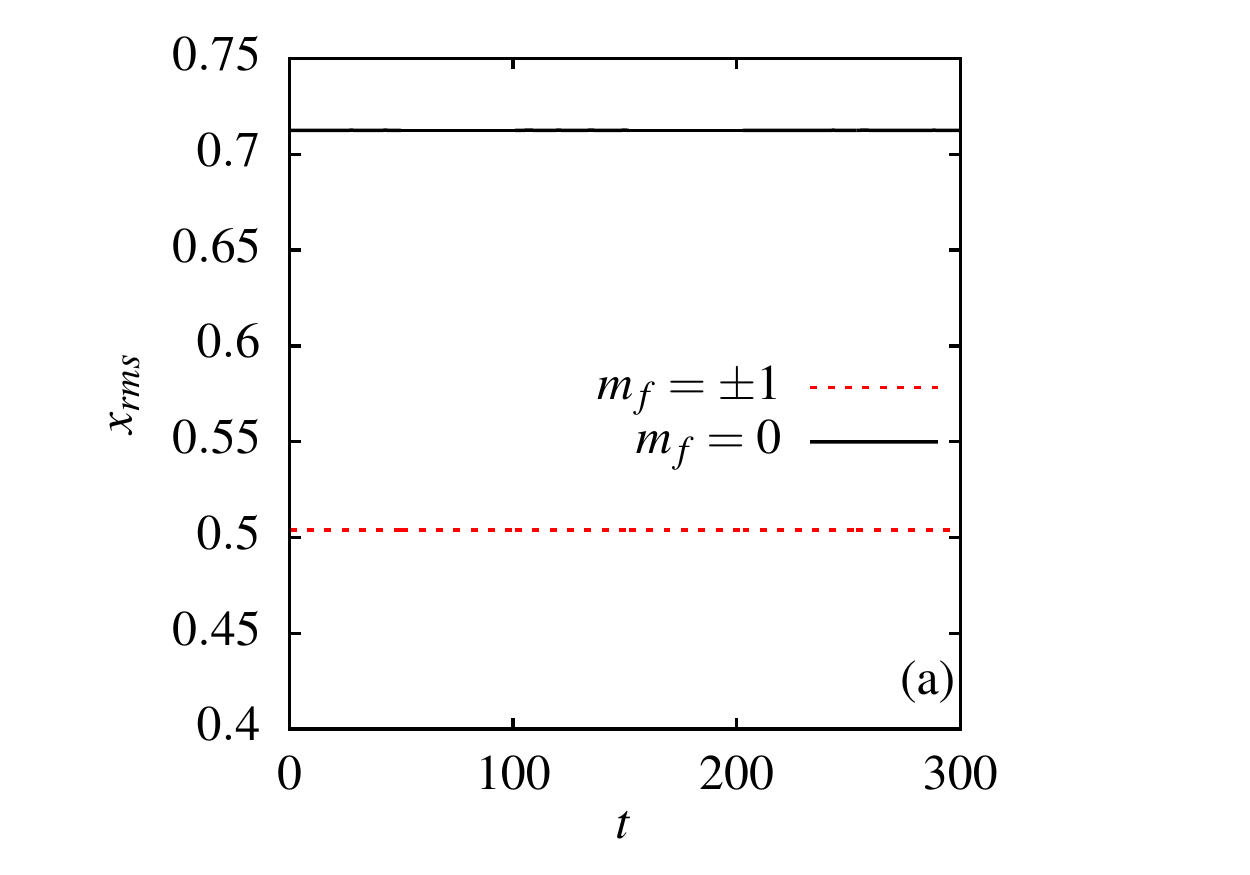}
	\includegraphics[trim = 3mm 0mm 0mm 0mm,clip, width=0.535\linewidth,clip]
	{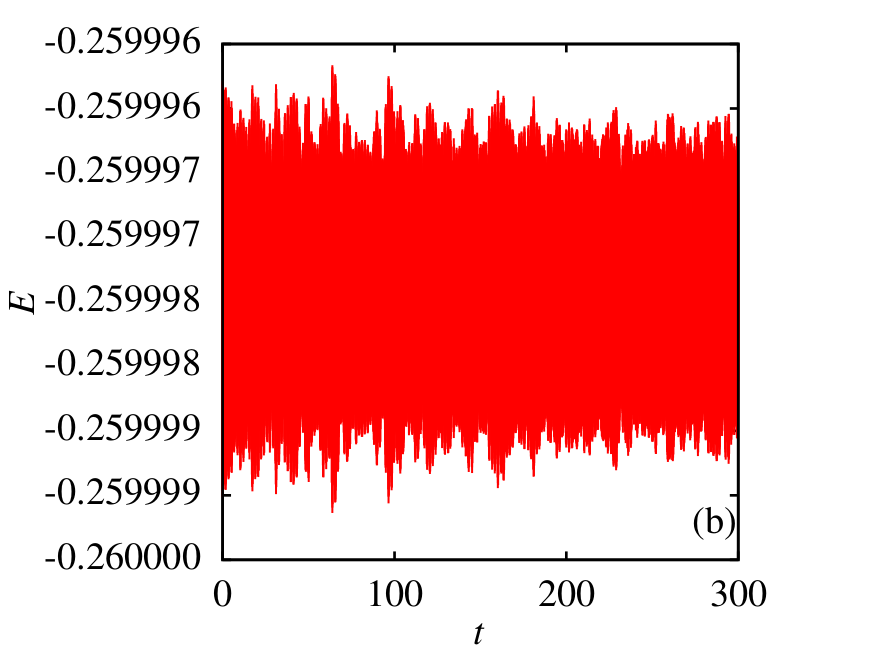}
	\caption{ (Color online) (a) Root mean square sizes of the three components 
         of $^{87}$Rb with $c_0 = -1.5, c_2 = -0.3$ and $\gamma_x = 0.5$ in the 
         absence of trap as a function of time. (b) Energy of the vector soliton
         as function of time.}
	\label{fig6}
\end{figure}

To check the stationary nature of the solutions one can evolve these
solutions using real-time propagation. As an example, we consider the
real-time evolution of a self-trapped solution of q1D $^{87}$Rb condensate 
with $c_0 = -1.5, c_2 = -0.3$ and $\gamma_x = 0.5$, which has $E = -0.2600$
as indicated in table \ref{table5}. The rms size of the three components of 
the vector soliton as a function of time is shown in Fig. \ref{fig6}(a). 
Similarly, energy $E$ as a function of $t$ is shown in Fig. \ref{fig6}(b)
which agrees with reported value of $-0.2600$ at all the times. 
All the results reported in this work confirm with this real-time evolution
check.

\subsection{Results for q2D and 3D spin-1 BECs}
Here we first consider $10^4$ atoms of $^{87}$Rb with 
$(a_0,a_2) = (5.387,5.313)$ nm in a q2D trap with $\alpha_x = \alpha_y = 1$,
$\alpha_z = 20$, $\omega_x/(2\pi) = 20$Hz. Secondly, we consider $10^4$ atoms 
of $^{23}$Na with $(a_0,a_2) = (2.646,2.911)$ nm in a q2D trap with same 
trapping frequencies as that for $^{87}$Rb. This leads to 
$(c_0,c_2) = (496.4428,-2.2942)$ and $(134.9838,4.2242)$ for $^{87}$Rb 
and $^{23}$Na, respectively. The ground state energies
(in the units $\hbar\omega_x$) for various magnetizations are given in table
\ref{2dtable}.
\begin{table}[H]
\caption{Ground state energies for $^{87}$Rb and $^{23}$Na q2D BECs 
obtained in the present work with $\Delta x = 0.05$, $\Delta y = 0.05$  
and $\Delta t = 0.1 \Delta x \Delta y$/2 for the various values of 
magnetization $\cal M$. $10^4$ atoms of each species were considered in
trap with $\alpha_x = \alpha_y = 1$, $\alpha_z = 20$, $\omega_x/(2\pi) = 20$Hz.
Together with scattering lengths $(a_0,a_2)$, these parameters define $c_0$
and $c_2$ as per Eq. (\ref{interaction2d}).}
\begin{center}
\begin{tabular}{| c | c | c |}
\hline
\multicolumn{1}{ |c| }{}&\multicolumn{1}{|c|}{$^{87}$Rb}&
\multicolumn{1}{ |c| }{$^{23}$Na}\\
\hline
\multicolumn{1}{ |c| }{$\cal M$}&\multicolumn{1}{|c|}{Energy}&
\multicolumn{1}{ |c| }{Energy}\\
\hline
0    & 8.4629 &  4.5355  \\ \hline
0.1  & 8.4629 &  4.5361 \\ \hline
0.2  & 8.4629 &  4.5380  \\ \hline
0.3  & 8.4629 &  4.5412 \\ \hline
0.4  & 8.4629 &  4.5457 \\ \hline
0.5  & 8.4629 &  4.5515 \\ \hline
0.6  & 8.4629 &  4.5586 \\ \hline
0.7  & 8.4629 &  4.5671 \\ \hline
0.8  & 8.4629 &  4.5771 \\ \hline
0.9  & 8.4629 &  4.5885  \\ \hline
\end{tabular}	
\label{2dtable}
\end{center}
\end{table}

\begin{figure}[H]
\begin{center}
\begin{tabular}{lll}
\includegraphics[trim = 0.4cm 0.4cm 0.4cm 0.4cm, clip,width=0.35\linewidth,
clip]{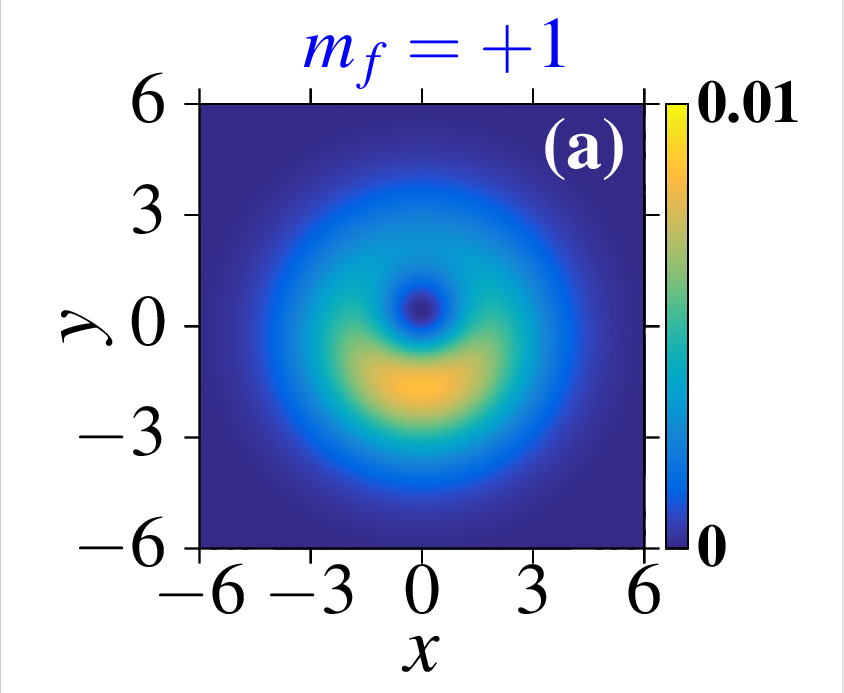}
\includegraphics[trim = 0.4cm 0.4cm 0.4cm 0.4cm, clip,width=0.35\linewidth,
clip]{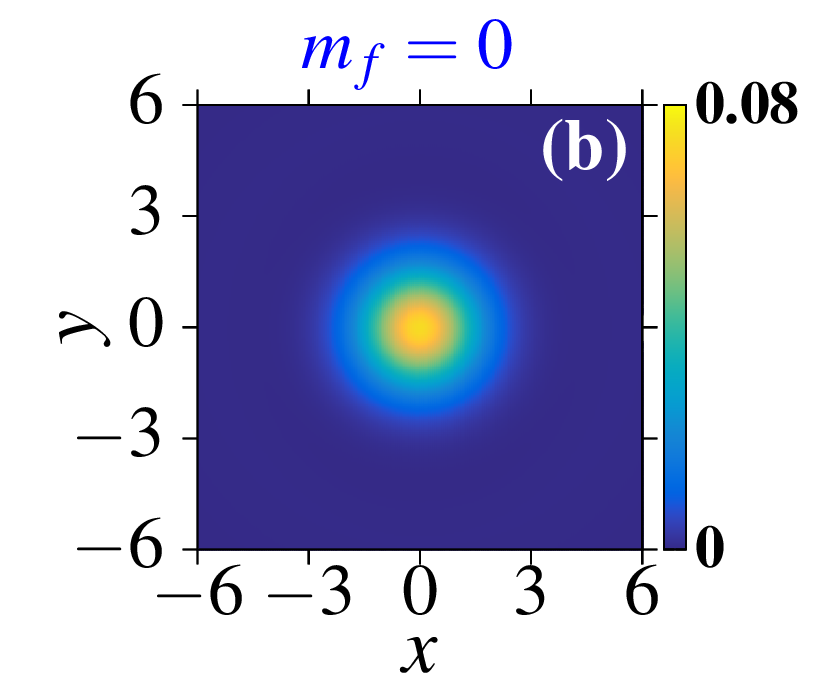}
\includegraphics[trim = 0.4cm 0.4cm 0.4cm 0.4cm, clip,width=0.35\linewidth,
clip]{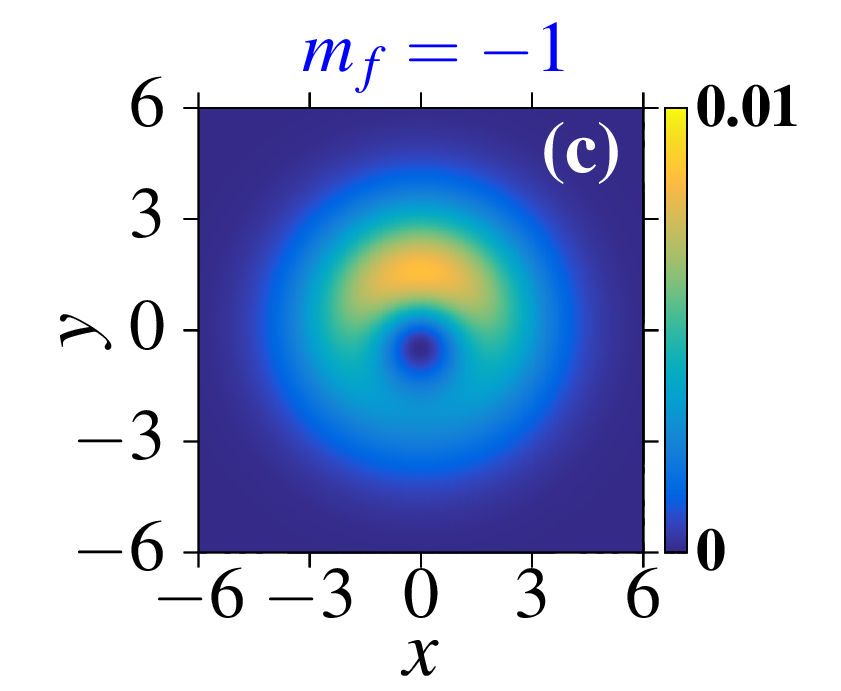}\\
\includegraphics[trim = 0.4cm 0.4cm 0.4cm 0.4cm, clip,width=0.35\linewidth,
clip]{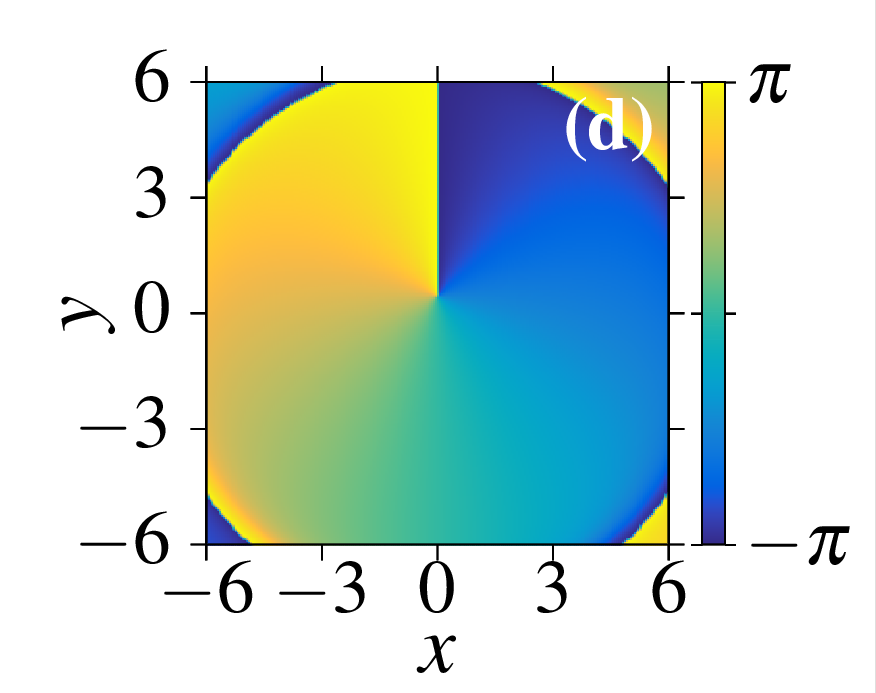}
\includegraphics[trim = 0.4cm 0.4cm 0.4cm 0.4cm, clip,width=0.35\linewidth,
clip]{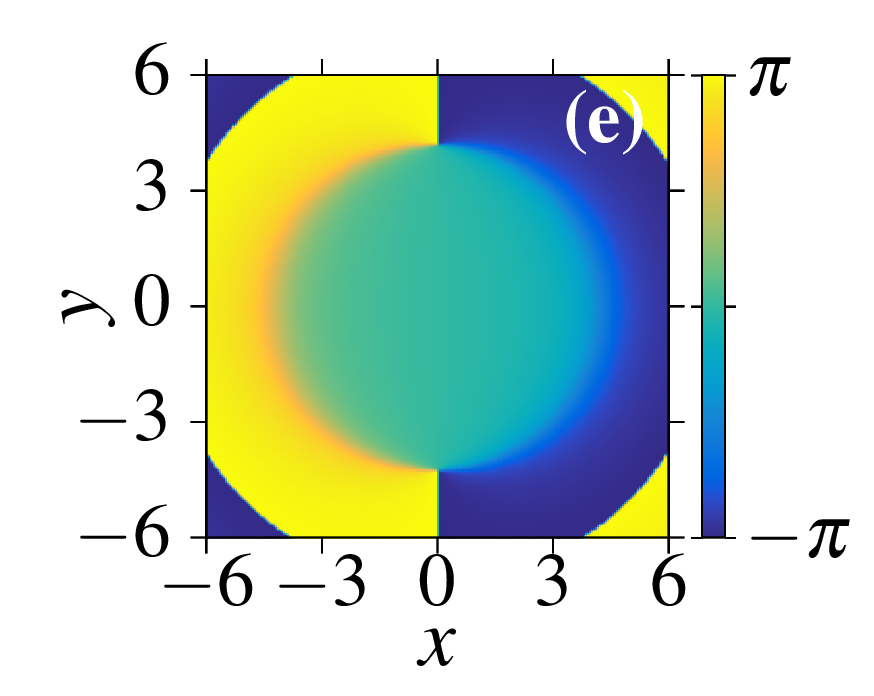}
\includegraphics[trim = 0.4cm 0.4cm 0.4cm 0.4cm, clip,width=0.35\linewidth,
clip]{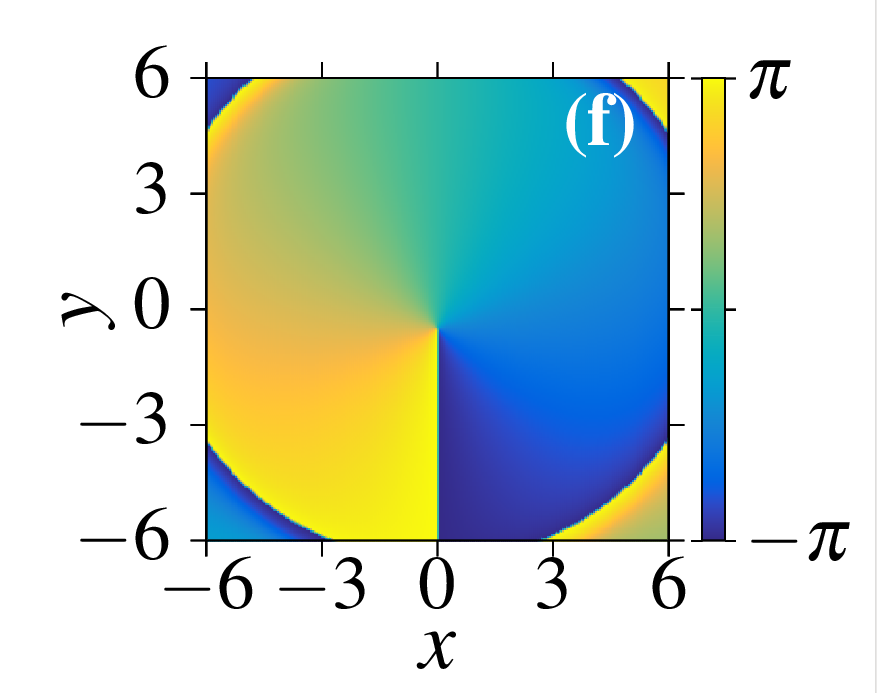}
\end{tabular}
\caption{The 2D contour plot of densities of (a) $m_f = +1$ , 
(b) $m_f = 0$, and (c) $m_f = -1$ components of an asymmetric vortex-bright 
soliton with $c_0 = -4$, $c_1 = -0.6$ and $\gamma_x=\gamma_y = 0.5$. The corresponding 
phases are shown in (d) for $m_f = +1$, (e) for $m_f = 0$ and (f) for 
$m_f = -1$ components.}
\label{fig4}
\end{center}
\end{figure}

\begin{figure}[H]
\begin{center}
\begin{tabular}{lll}
\includegraphics[trim = 0.4cm 0.4cm 0.4cm 0.4cm, clip,width=0.35\linewidth,
clip]{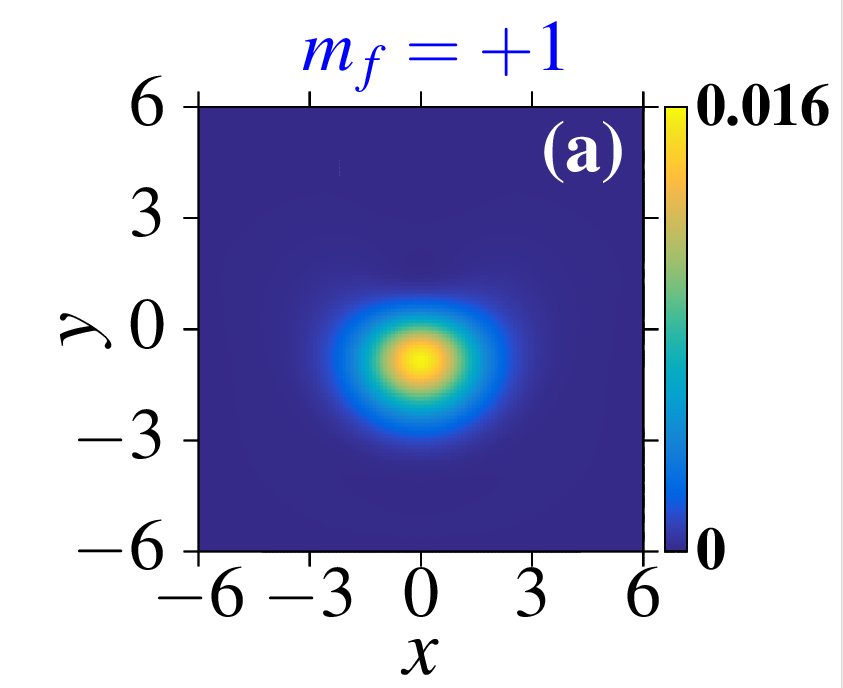}
\includegraphics[trim = 0.4cm 0.4cm 0.4cm 0.4cm, clip,width=0.35\linewidth,
clip]{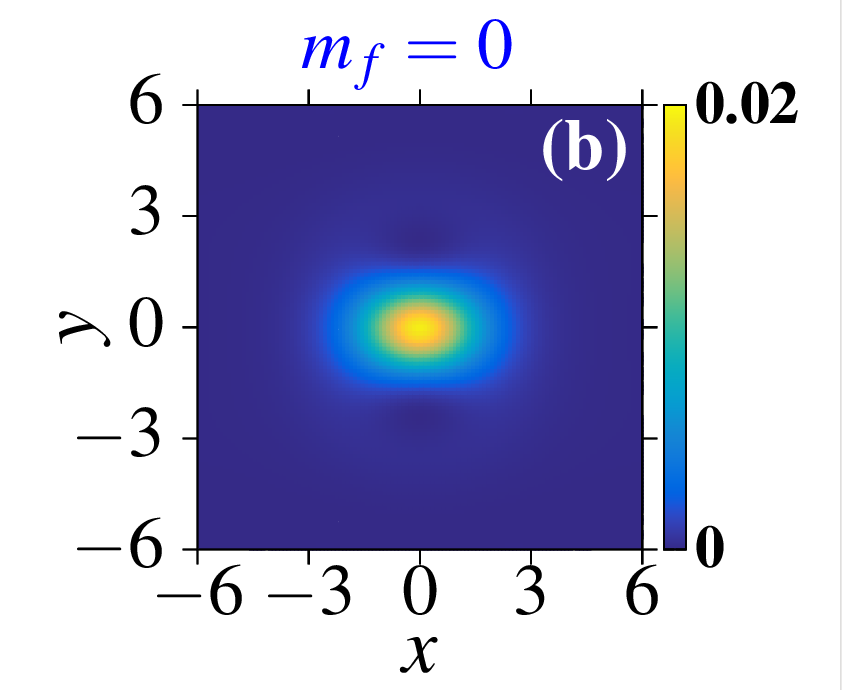}
\includegraphics[trim = 0.4cm 0.4cm 0.4cm 0.4cm, clip,width=0.35\linewidth,
clip]{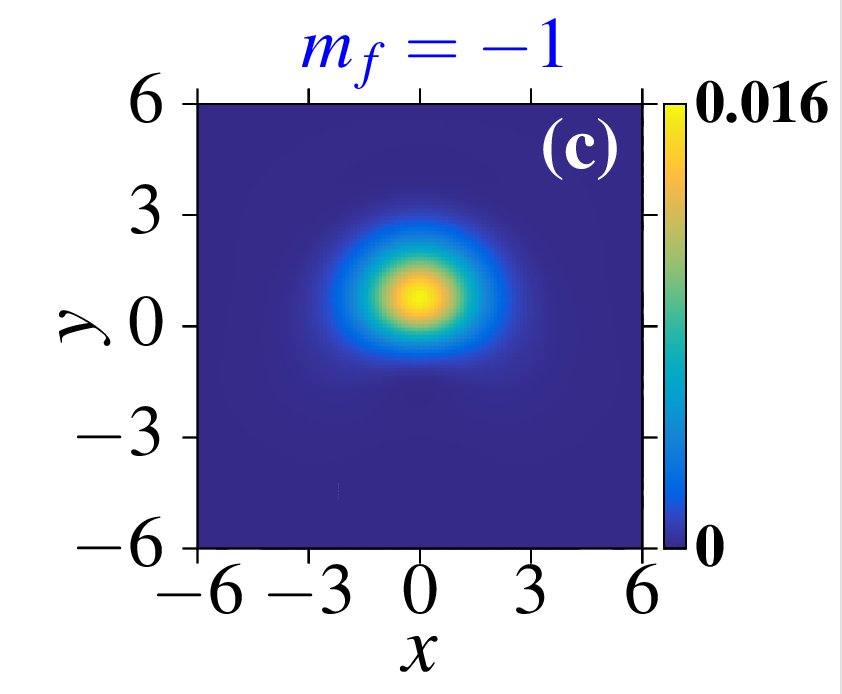}\\
\includegraphics[trim = 0.4cm 0.4cm 0.4cm 0.4cm, clip,width=0.35\linewidth,
clip]{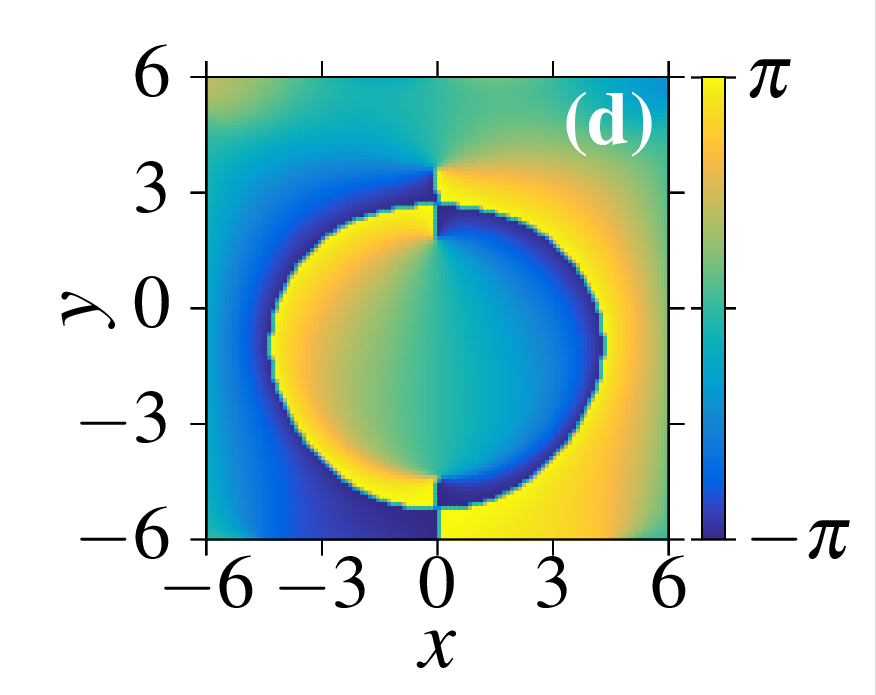}
\includegraphics[trim = 0.4cm 0.4cm 0.4cm 0.4cm, clip,width=0.35\linewidth,
clip]{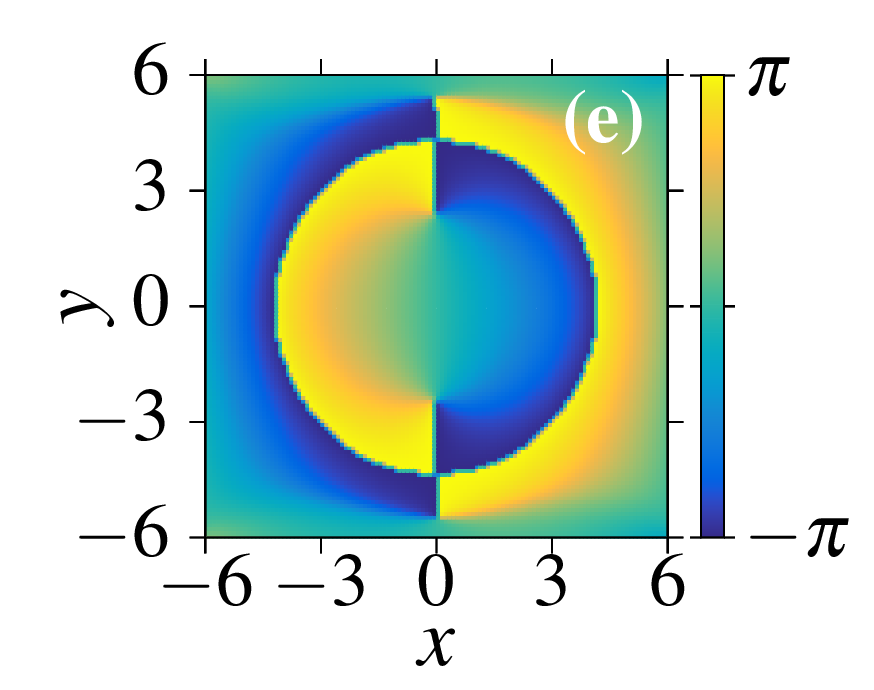}
\includegraphics[trim = 0.4cm 0.4cm 0.4cm 0.4cm, clip,width=0.35\linewidth,
clip]{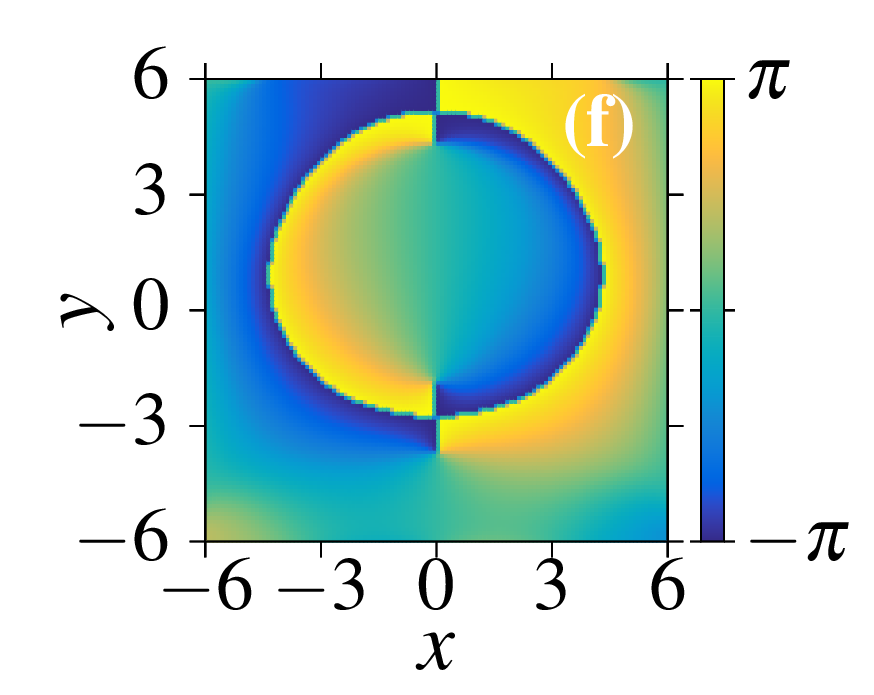}
\end{tabular}
\caption{The 2D contour plots of densities of components in $z=0$ plane for 
(a) $m_f = +1$, (b) $m_f = 0$, and (c) $m_f = -1$ of an asymmetric 
vortex-bright soliton with $c_0 = -10,$ $c_1 = -1$ and $\gamma_x=\gamma_y=\gamma_z =1$. The 
corresponding phases are shown in (d) for $m_f = +1$, (e) for $m_f = 0$, 
and (f) for $m_f = -1$ components.
}
\label{fig5}
\end{center}
\end{figure}

For q2D case, we also consider $c_0 = -4, c_1 = -0.6$ with 
$\gamma_x=\gamma_y = 0.5$, i.e isotropic SO coupling, in the absence 
of trapping. The ground state in this case is a self-trapped vortex-bright 
soliton as is shown in Fig. \ref{fig4}. The ground state solution corresponds 
to an asymmetric antivortex and vortex in the $m_f = +1$ and $m_f  = -1$ 
components, respectively as is illustrated in Fig. \ref{fig4}(d)-(f) 
\cite{gautam2017vortex}.
Similarly in the 3D case, we consider $c_0 = -10, c_1 = 0.1$ with 
$\gamma_x = \gamma_y = \gamma_z = 1$ in the absence of trapping. Again, the 
ground state solution in this case is a self-trapped vortex-bright soliton. 
To illustrate this vortex-bright soliton, we plot the two-dimensional contour 
densities and corresponding phase profiles in $z=0$ plane in Fig. \ref{fig5}. 
These results are in agreement with \cite{gautam2018three}.

\section{Summary}
\label{section-6}
We have discussed a time-splitting Fourier spectral method to solve the
mean-field model of spin-1 BECs with anisotropic spin-orbit coupling.
The time-splitting coupled with spectral method allows one to deal with
non-linear and SO coupling terms very precisely. The numerical scheme has been 
implemented via three FORTRAN 90/95 codes, which are OpenMP parallelized, 
for quasi-one, quasi-two and three-dimensional spin-1 BECs. 
We have provided the results for execution time, speedup, and
efficiency as a function of number of threads for the three codes.
The numerical results obtained 
with the three codes are in very good agreement with previous results without
SO coupling from the literature. The model of SO coupling is quite general 
enough to allow the users of the codes to simulate a variety of SO couplings 
considered in the literature which include Rashba SO coupling 
(isotropic or anisotropic), Dresselhaus SO coupling (isotropic or anisotropic) 
or their mixture. With the recent spur in the studies on SO coupled 
spinor BECs, the present numerical scheme along with the codes 
could be quite useful to the researchers exploring this field.
The spectral method used in manuscript can be extended to solve the 
Stochastic projected coupled Gross-Pitaevskii equations for spin-1 BECs
on one hand (here the implementation of the projection is quite natural
in Fourier space) or simulate rotating Spin-1 BECs with or without SO coupling.
These two directions may be explored in future projects.

\section*{Acknowledgments}
AR acknowledges support from Provincia Autonoma di Trento. 
SG thanks the Science \& Engineering Research Board, 
Department of Science and Technology, Government of India (Project: 
ECR/2017/001436) and Indian Institute of Technology, 
Ropar (ISIRD Project: 9-256/2016/IITRPR/823) for support. 
SG acknowledges the useful discussions with Prof. S. K. Adhikari of 
Instituto de F\'{\i}sica Te\'orica, Universidade Estadual Paulista, 
S\~ao Paulo.


\begin{thebibliography}{10}
	\bibitem{anderson1995observation}
	M.H. Anderson, J.R. Ensher, M.R. Matthews, C.E. Wieman, and 
	E.A. Cornell, 
	Science 269 (1995) 198;
	K.B. Davis, M.-O. Mewes, M.R. Andrews, N.J. van Druten, D.S. Durfee, 
	D.M. Kurn, and W. Ketterle, 
	Phys. Rev. Lett.  75 (1995) 3969;
	C.C. Bradley, C.A. Sackett, J.J. Tollett, and R.G. Hulet,
	Phys. Rev. Lett.  75 (1995) 1687.
	
	\bibitem{stamper1998optical}
	D.M. Stamper-Kurn, M.R. Andrews, A.P. Chikkatur, S. Inouye, H.-J. 
	Miesner, J. Stenger, and W. Ketterle,
	Phys. Rev. Lett. 80 (1998) 2027.
	
	\bibitem{stenger1998spin1}
	J. Stenger, S. Inouye, D.M. Stamper-Kurn, H.-J. Miesner, A.P. Chikkatur, 
	and W. Ketterle, Nature 396 (1998) 345; 
	M. D. Barrett, J. A. Sauer, and M. S. Chapman,
	Phys. Rev. Lett.  87 (2001) 010404.
	A.T. Black, E. Gomez, L.D. Turner, S. Jung, and P.D. Lett,
	Phys. Rev. Lett. 99 (2007) 070403;
	M.-S. Chang, C.D. Hamley, M.D. Barrett, J.A. Sauer, K.M. Fortier, 
	W. Zhang, L. You, and M.S. Chapman,
	Phys. Rev. Lett. 92 (2004) 140403;
	A. G\"orlitz, T.L. Gustavson, A.E. Leanhardt, R. L\"ow, A.P. 
	Chikkatur, S. Gupta, S. Inouye, D.E. Pritchard, and W. Ketterle,
	Phys. Rev. Lett. 90 (2003) 090401;
	H. Schmaljohann, M. Erhard, J. Kronj\"ager, M. Kottke, S. van Staa, 
	L. Cacciapuoti, J.J. Arlt, K. Bongs, and K. Sengstock,
	Phys. Rev. Lett. 92 (2004) 040402;
	T. Kuwamoto, K. Araki, T. Eno, and T. Hirano,
	Phys. Review A 69 (2004) 063604;
	B. Pasquiou, E. Mar\'echal, G. Bismut, P. Pedri, L. Vernac, O. Gorceix,
	and B. Laburthe-Tolra,
	Phys. Rev. Lett. 106 (2011) 255303;
	D.M. Stamper-Kurn and M. Ueda,
	Rev. Mod. Phys. 85 (2013) 1191.
    	
	\bibitem{osterloh2005cold}
	K. Osterloh, M. Baig, L. Santos, P. Zoller, and M. Lewenstein,
	Phys. Rev. Lett. 95 (2005) 010403;
	J. Ruseckas, G. Juzeli{\=u}nas, P. {\"O}hberg, and M. Fleischhauer,
	Phys. Rev. Lett. 95 (2005) 010404;
	J. Dalibard, F. Gerbier, G. Juzeli{\=u}nas, and P. {\"O}hberg,
	Rev. Mod. Phys. 83 (2011) 1523;
	N. Goldman, G. Juzeli{\=u}nas, P. {\"O}hberg, and I.B. Spielman,
	Rep. Prog. Phys. 77 (2014) 126401.
	\bibitem{lin2011spin}
	Y.-J. Lin, K. Jim{\'e}nez-Garc{\'\i}a, and I.B. Spielman,
	Nature 471 (2011) 83.
	
	\bibitem{bychkov1984oscillatory}
	Y.A. Bychkov and E.I. Rashba,
	J. Phys. C: Solid state physics 17 (1984) 6039.
	
	\bibitem{dresselhaus1955spin}
	G. Dresselhaus,
	Phys. Rev. 100 (1955) 580.
	\bibitem{aidelsburger2011experimental}
	M. Aidelsburger, M. Atala, S. Nascimb\'ene, S. Trotzky, Y.-A. Chen, 
	and I. Bloch,
	Phys. Rev. Lett. 107 (2011) 255301;
	Z. Fu, P. Wang, S. Chai, L. Huang, and J. Zhang ,
	Phys. Rev. A 84 (2011) 043609;
	J.-Y. Zhang, S.-C. Ji, Z. Chen, L. Zhang, Z.-D. Du, B. Yan, G.-S. Pan, 
	B. Zhao, Y.-J. Deng, H. Zhai, S. Chen, and J.-W Pan,
	Phys. Rev. Lett. 109 (2012) 115301;
	C. Qu, C. Hamner, M. Gong, C. Zhang, and P. Engels,
	Phys. Rev. A 88 (2013) 021604.
	
	\bibitem{wang2010spin}
	C. Wang, C. Gao, C.-M. Jian, and H. Zhai,
	Phys. Rev. Lett. 105 (2010) 160403.
	
	\bibitem{zhai2012spin}
	H. Zhai,
	Int. J. Mod. Phys. B 26 (2012) 1230001.
		
	\bibitem{sinha2011trapped}
	S. Sinha, R. Nath, and L. Santos,
	Phys. Rev. Lett. 107 (2011) 270401;	
	H. Hu, B. Ramachandhran, H. Pu, and X.-J. Liu,
	Phys. Rev. Lett. 108 (2012) 010402;
	Y. Xu, Y. Zhang, and B. Wu,
	Phys. Review A 87 (2013) 013614;
	L. Salasnich and B. A. Malomed,
	Phys. Review A 87 (2013) 063625;
	Salasnich, L., Cardoso, W.~B., and Malomed, B.~A.,
	Phys. Rev. A 90 (2014) 033629;
	S. Cao, C.-J. Shan, D.-W. Zhang, X. Qin, and J. Xu,
	JOSA B 32 (2015) 201;
	H. Sakaguchi, B. Li, and B.A. Malomed,
	Phys. Rev. E 89 (2014) 032920;
	H. Sakaguchi and B.A. Malomed,
	Phys. Rev. E 90 (2014) 062922;
	Y.-K. Liu and S.-J. Yang,
	Euro Phys. Lett. 108 (2014) 30004;
	T.-L. Ho and S. Zhang,
	Phys. Rev. Lett. 107 (2011) 150403;
	Z.-F. Xu, L. You, and M. Ueda,
	Phys. Rev. A 87 (2013) 063634
	
	\bibitem{galitski2013spin}
	V. Galitski and I.B. Spielman,
	Nature 494 (2013) 49.
	
	\bibitem{hasan2010colloquium}
	C.L. Kane and E.J. Mele,
	Phys. Rev. Lett. 95 (2005) 146802;
	B.A. Bernevig, T.L. Hughes, and S.C. Zhang,
	Science 314 (2006) 1757;
	D. Hsieh, D. Qian, L. Wray, Y. Xia, Y. S. Hor, R.J. Cava, and 
	M.Z. Hasan,
	Nature 452 (2008) 970;
	M.Z. Hasan, and C.L. Kane,
	Rev. Mod. Phys. 82 (2010) 3045;
	X.L. Qi and S.-C. Zhang,
	Rev. Mod. Phys. 83 (2011) 1057.
	
	\bibitem{koralek2009emergence}
	J.D. Koralek, C.P. Weber, J. Orenstein, B.A. Bernevig, 
	S.-C. Zhang, S. Mack, and D.D. Awschalom 
	Nature 458 (2009) 610.
	\bibitem{avsar2014spin}
	A. Avsar {\em et al.},
	Nature Comm. 5 (2014) 4875;
	Z. Wang, C. Tang, R. Sachs, Y. Barlas, and J. Shi,
	Phys. Rev. Lett. 114 (2015) 016603.
	
	\bibitem{sau2010generic}
	J.D. Sau, R.M. Lutchyn, S. Tewari, and S.D. Sarma,
	Phys. Rev. Lett. 104 (2010) 040502.
	\bibitem{campbell2016magnetic}
	D.L. Campbell, R.M. Price, A. Putra, A. Vald\'es-Curiel, 
	D. Trypogeorgos and I.B. Spielman ,
	Nature communications 7 (2016) 10897.
	
	\bibitem{liu2019skyrmions}
	Y.-K. Liu, G.-H. Yang, L.-L. Xu, and S.-J. Yang,
	Ann. Phys. 405 (2019) 289;
	L. Zhang, Y. Ke, and C. Lee,
	Phys. Rev. B 100 (2019) 224420;
	Y.V. Kartashov V.V. and Konotop,
	Phys. Rev. Lett. 118 (2017) 190401.

	\bibitem{li2017stripe}
	J.-R. Li, J. Lee, W. Huang, S. Burchesky, B. Shteynas, F.C. Top, 
    A.O. Jamison, and W. Ketterle,
	Nature 543 (2017) 91;
	J. Li, W. Huang, B. Shteynas, S. Burchesky, F.C. Top, E. Su, J. Lee, 
	A.O. Jamison, and W. Ketterle
	Phys. Rev. Lett. 117 (2016) 185301;
	S. Zhang, and G.-B. Jo,
	Journal of Physics and Chemistry of Solids 
	128 (2018) 75.	

    \bibitem{zwartsenberg20}
    Zwartsenberg, B. \emph{et al.},
    Nature Physics, https://doi.org/10.1038/s41567-019-0750-y 

	\bibitem{ho1998spinor}
	T.-L. Ho,
	Phys. Rev. Lett. 81 (1998) 742.
	\bibitem{ohmi1998bose}
	T. Ohmi and K. Machida,
	Journal of the Physical Society of Japan {\bf 67} (1998) 1822.
	\bibitem{wang2007time}
	H. Wang,
	Int. J. Comp. Math. {\bf 84} (2007) 925.
	\bibitem{bao2008computing}
	W. Bao and F.Y. Lim, F.~Y.,
	SIAM Journal on Scientific Computing 30 (2008) 1925.
	
	\bibitem{bao2013efficient}
	W. Bao, I.-L. Chern, and Y. Zhang,
	J. Comp. Phys. 253 (2013) 189.
	\bibitem{ruprecht1995time}
	P. Ruprecht, M. Holland, K. Burnett, and M. Edwards,
	Phys. Rev. A 51 (1995) 4704;
	M. Edwards and K. Burnett,
	Phys. Rev. A 51 (1995) 1382.
	R. Dodd,
	Journal of Research of the National Institute of Standards and
	Technology 101 (1996) 545;
	W. Bao and W. Tang,
	J. Comp. Phys. 187 (2003) 230;
    W. Bao and Y. Cai,
    SIAM Journal on Numerical Analysis 50 (2012) 492;
    R.P. Tiwari and A. Shukla,
    Comp. Phys. Comm. 174 (2006) 966;
    W. Bao, W. and Y. Cai,
    Mathematics of Computation 82 (2013) 99.
	
	\bibitem{chiofalo2000ground}
	M.L. Chiofalo, S. Succi, and M. Tosi,
	Phys. Rev. E 62 (2000) 7438
	
	\bibitem{bao2004computing}
	 W. Bao and Q. Du,
	SIAM Journal on Scientific Computing 25 (2004) 1674
	
	\bibitem{antoine2013computational}
	X. Antoine, W. Bao, and C. Besse,
	Comp. Phy. Comm. 184 (2013) 2621.
	
	\bibitem{muruganandam2009fortran}
	P. Muruganandam and S.K. Adhikari,
	Comp. Phys. Comm. 180 (2009) 1888
	
	\bibitem{bao2003numerical}
	 W. Bao, D. Jaksch, and P.A. Markowich,
	J. Comp. Phys. 187 (2003) 318.
	
	\bibitem{chang2005gauss}
	S.-M. Chang, W.-W. Lin, and S.-F. Shieh,
	J. Comp. Phys. 202 (2005) 367;
	W. Bao and J. Shen,
	SIAM Journal on Scientific Computing 26 (2005) 2010.
	
	\bibitem{wang2014projection}
	H. Wang and Z. Xu,
	Comp. Phys. Comm. 185 (2014) 2803;
	H. Wang, J. Comp. Phys. 274 (2014) 473.
	\bibitem{bao2002time}
	W. Bao, S. Jin, P.A. Markowich,
	J. Comp. Phys. 175 (2002) 487;
	W. Bao, S. Jin, P.A. Markowich, 
	SIAM Journal on Scientific Computing, 25 (2003) 27.

    \bibitem{kumar_19}
     L.E. Young-S., P. Muruganandam, S.K. Adhikari, 
	V. Loncar, D. Vudragovic, A. Balaz
	Comput. Phys. Commun. 220 (2017) 503;
	V. Loncar, L.E. Young-S., S. Skrbic, 
	P. Muruganandam, S.K. Adhikari, Antun Balaz
	Comput. Phys. Commun. 209 (2016) 190;
	L.E. Young-S., D. Vudragovic, P. Muruganandam, 
	S.K. Adhikari, A. Balaz
	Comput. Phys. Commun. 204 (2016) 209;
	B. Sataric, V. Slavnic, A. Belic,
	 A. Balaz, P. Muruganandam, S.K. Adhikari
	Comput. Phys. Commun. 200 (2016) 411;
    V. Loncar, A. Balaz, A. Bogojevic,
     S. Skrbic, P. Muruganandam, S.K. Adhikari
    Comput. Phys. Commun. 200 (2016) 406;
	D. Vudragovic, I. Vidanovic, A. Balaz,
    P. Muruganandam, S.K. Adhikari
	Comput. Phys. Commun. 183 (2012) 2021;
	X. Antoine and R. Duboscq
	Comput. Phys. Commun. 185 (2014) 2969;
	X. Antoine and R. Duboscq
	Comput. Phys. Commun. 193 (2015) 95;
	{\v Z}. Marojevi{\'c} and E. G{\" o}kl{\" u} and 
	Claus L{\"a}mmerzahl Comput. Phys. Commun. 202 (2016) 216

	\bibitem{ville_18}
        J.L. Ville R. Saint-Jalm, {\'E}. Le Cerf, M. Aidelsburger, 
        S. Nascimb{\`e}ne, J. Dalibard, and J. Beugnon,
        Phys. Rev. Lett. 121 (2018) 145301.
	
	
	
	\bibitem{campbell2016rashba}
	D.L. Campbell and I.B. Spielman,
	New Journal of Physics 18 (2016) 033035.
	
	\bibitem{gautam2018three}
	S. Gautam, and S.K. Adhikari,
	Phys. Rev. A 97 (2018) 013629.
	
	\bibitem{salasnich2002effective}
	L. Salasnich, A. Parola, and L. Reatto,
	Phys. Rev. A 65 (2002) 043614;
	L. Salasnich, A. Parola, and L. Reatto, Phys. Rev. A 72 (2005) 025602.
	
	\bibitem{gautam2017vortex}
	S. Gautam and S.K. Adhikari,
	Phys. Rev. A 95 (2017) 013608.
	
	\bibitem{gautam2015mobile}
	S. Gautam. and S.K. Adhikari,
	Laser Physics Letters 12 (2015) 045501.
	
	\bibitem{kumar2019c}
	R.K. Kumar, V. Lon{\v{c}}ar, P. Muruganandam, S.K. Adhikari, and
	A. Bala{\v{z}},
	Comp. Phys. Comm. 240 (2019) 74;
	R.K. Kumar, L.E. Young-S, D. Vudragovi\'c, A. Bala{\v{z}}, 
	P. Muruganandam, and S.K. Adhikari,
	Comp. Phys. Comm. 195 (2015) 117.
	
	\bibitem{FFTW}
	http://www.fftw.org/
        \bibitem{FORTRESS-DATA}
        P. Makkar, A. Roy, S. Gautam (2020), 
        “FORTRESS\_DATA”, Mendeley Data, 
        V1, doi: 10.17632/tct7vjh994.1	
\end{thebibliography}
\end{document}


\centering
{\Huge Electronic Appendix} \\~\\
 This contains the snapshots of the output files, file1\_im.dat/file1\_re.dat,  correspnding to different input parameters.\\
       This data is generated by using FORTRESS.
\begin{figure}
	{\bf (a)}
	\includegraphics[width=9.in]{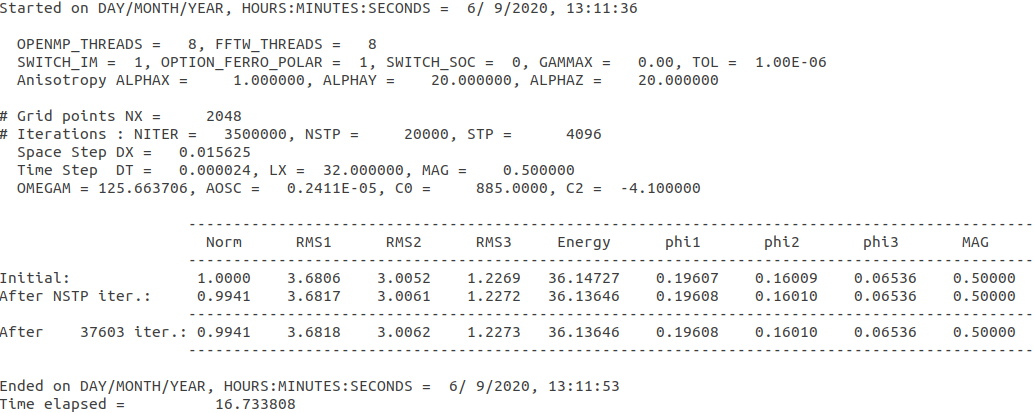}\\
 \rule{25cm}{1pt}	
	\\{\bf (b)}
 \includegraphics[width=9.in]{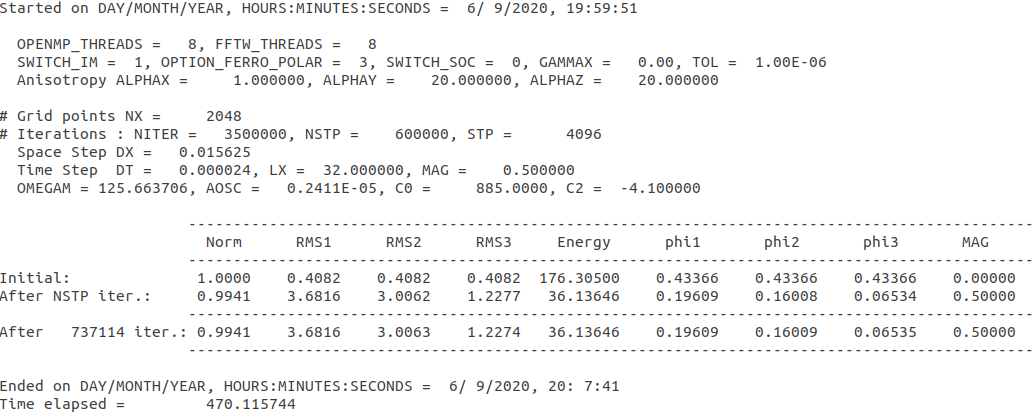}
 \caption{Contents of `file1\_im.dat' for $^{87}$Rb with
	(a) OPTION\_FERRO\_POLAR = 1 and (b) OPTION\_FERRO\_POLAR = 3 
obtained with two runs of `imretime\_spin1\_1D.f90'.}
\end{figure}

\begin{figure}
        {\bf (a)}
 \includegraphics[width=9.in]{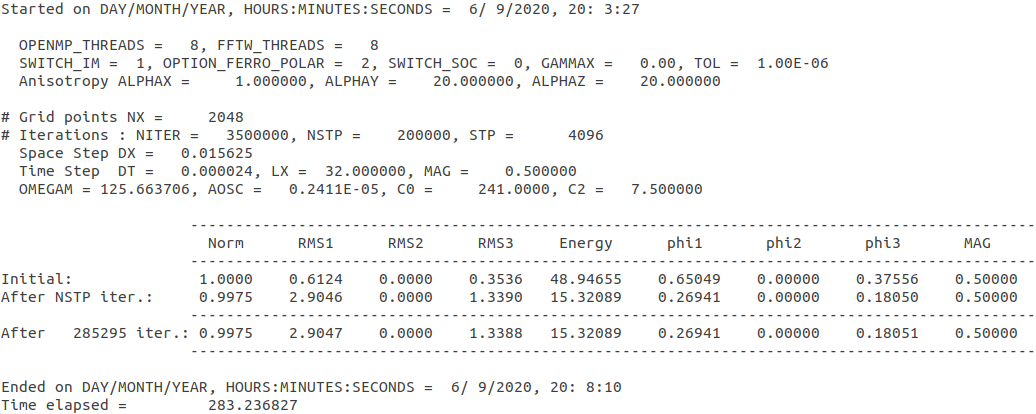}\\
 \rule{25cm}{1pt}
 \\{\bf (b)}
 \includegraphics[width=9.in]{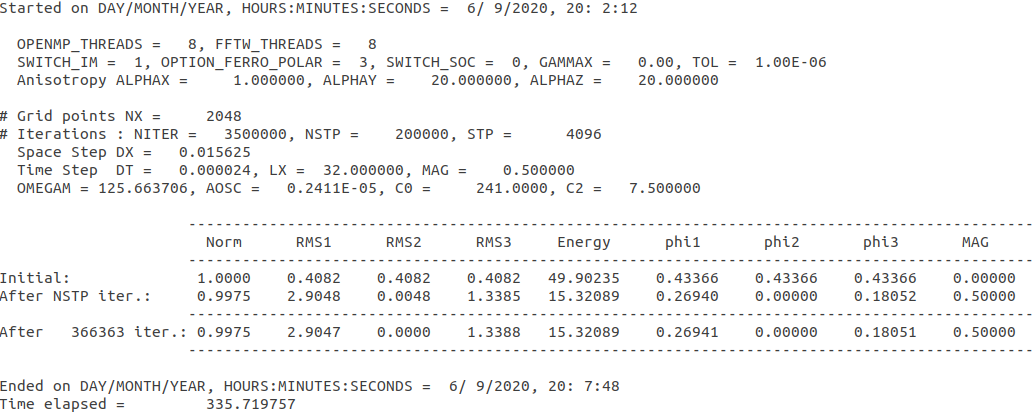}
 \caption{Contents of `file1\_im.dat' for $^{23}$Na with 
	(a) OPTION\_FERRO\_POLAR = 2 and (b) OPTION\_FERRO\_POLAR = 3  obtained with
`imretime\_spin1\_1D.f90'.}
\end{figure}

\begin{figure}
        {\bf (a)}
 \includegraphics[width=9.in]{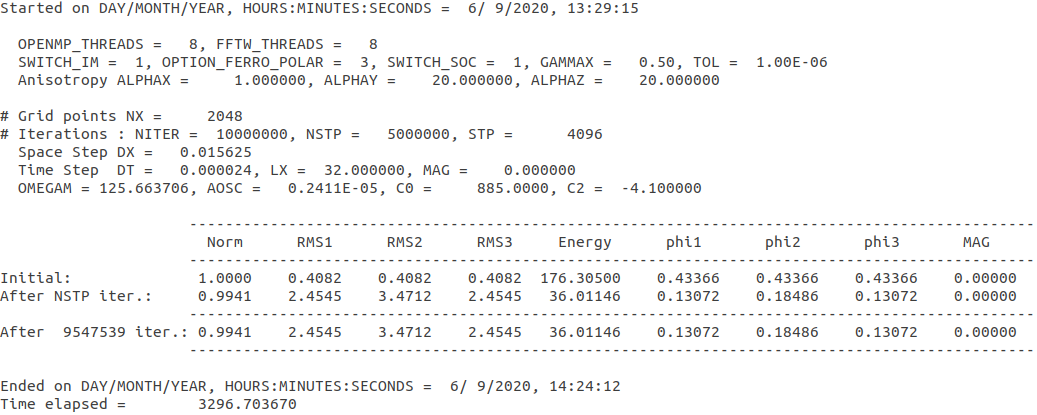}\\
\rule{25cm}{1pt} 
 \\{\bf (b)}
 \includegraphics[width=9.in]{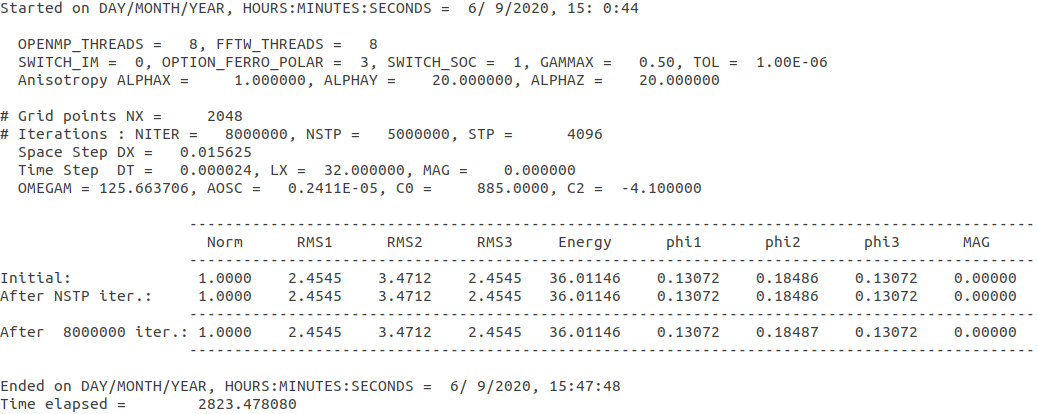}
	\caption{Contents of (a) `file1\_im.dat'  and (b) `file1\_re.dat'  for $^{87}$Rb with
non-zero SO coupling strength of $0.5$ obtained with `imretime\_spin1\_1D.f90'. The realtime simulation corresponds to
initial input wavefunctions obtained from the imaginary time run of the code.}
\end{figure}

\begin{figure}
        {\bf (a)}
 \includegraphics[width=9.in]{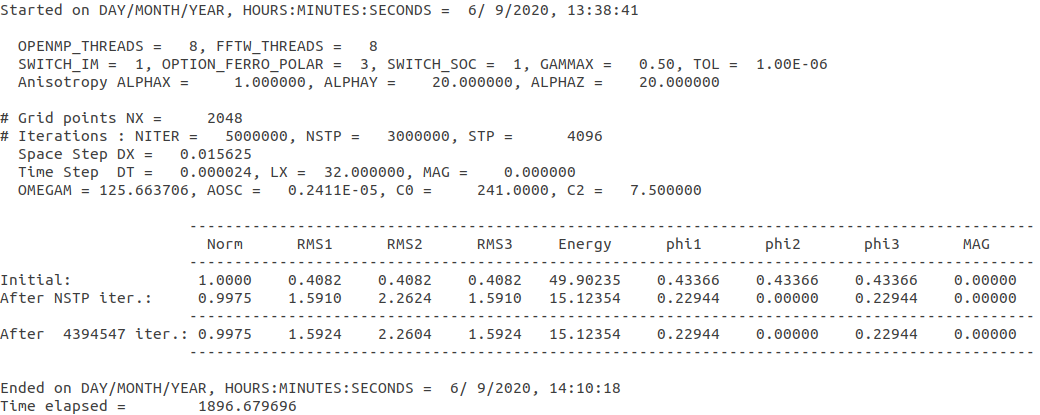}\\
 \rule{25cm}{1pt}
 \\{\bf (b)}
 \includegraphics[width=9.in]{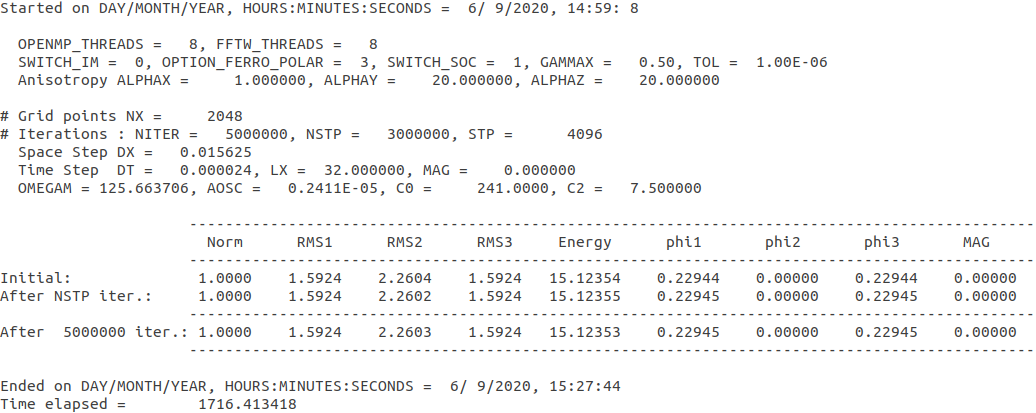}
	\caption{Contents of (a) `file1\_im.dat'  and (b) `file1\_re.dat' for $^{23}$Na with
non-zero SO coupling strength of $0.5$ obtained with `imretime\_spin1\_1D.f90'. The realtime simulations correspond to
initial input wavefunctions obtained from the imaginary time run of the code.}
\end{figure}

\begin{figure}
        {\bf (a)}
 \includegraphics[width=9.in]{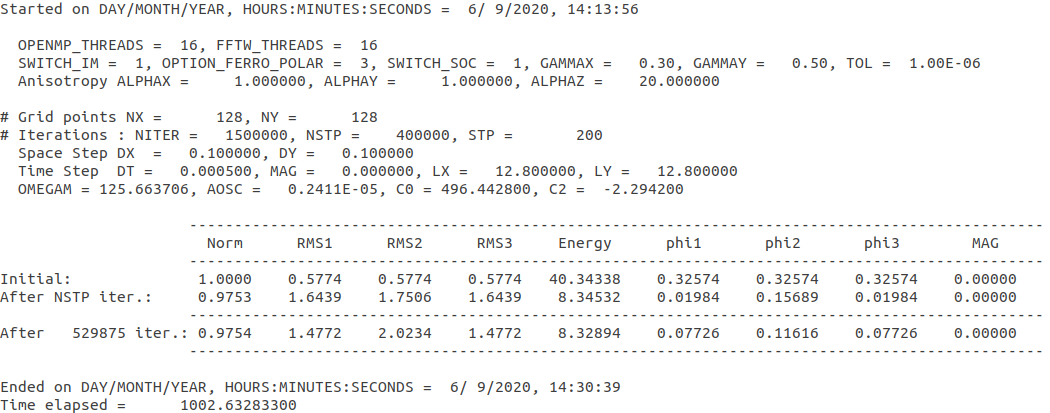}\\\rule{25cm}{1pt}
 \\{\bf (b)}
 \includegraphics[width=9.in]{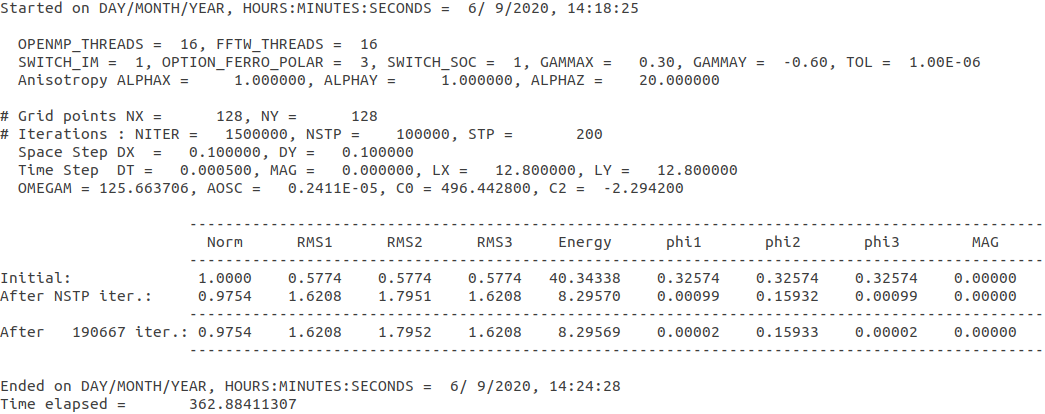}
        \caption{Contents of `file1\_im.dat' for quasi-2D $^{87}$Rb condensate with
	anisotropic SO coupling strengths of (a) $(\gamma_x,\gamma_y) = (0.3,0.5)$  and
(b) $(\gamma_x,\gamma_y) = (0.3,-0.6)$  obtained with `imretime\_spin1\_2D.f90'.}
\end{figure}

\begin{figure}
        {\bf (a)}
 \includegraphics[width=9.in]{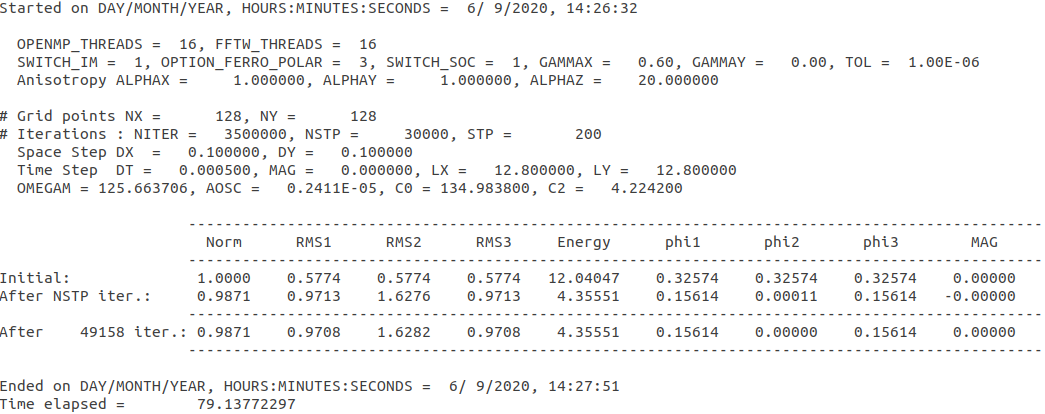}\\
 \rule{25cm}{1pt}\\{\bf (b)}
 \includegraphics[width=9.in]{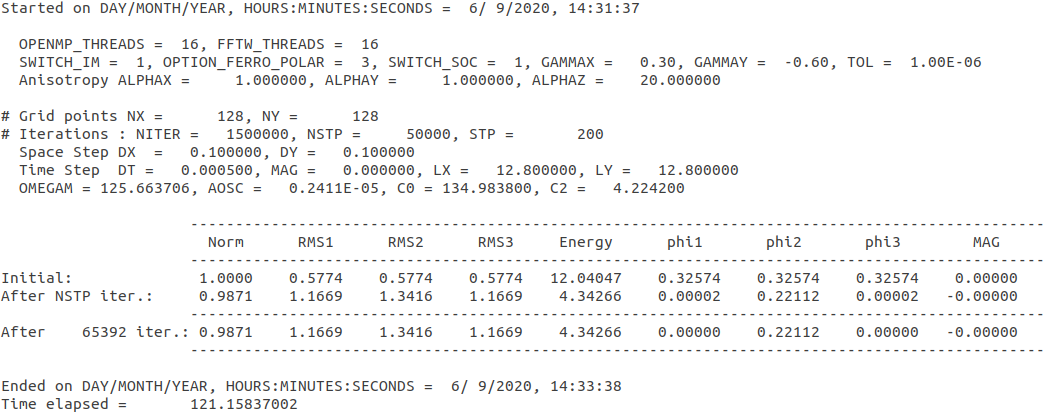}
        \caption{Contents of `file1\_im.dat' for quasi-2D $^{23}$Na condensate with
	anisotropic SO coupling strengths of (a) $(\gamma_x,\gamma_y) = (0.3,0.5)$  and
 (b) $(\gamma_x,\gamma_y) = (0.3,-0.6)$  obtained with `imretime\_spin1\_2D.f90'.}
\end{figure}

\begin{figure}
        {\bf (a)}
 \includegraphics[width=9.in]{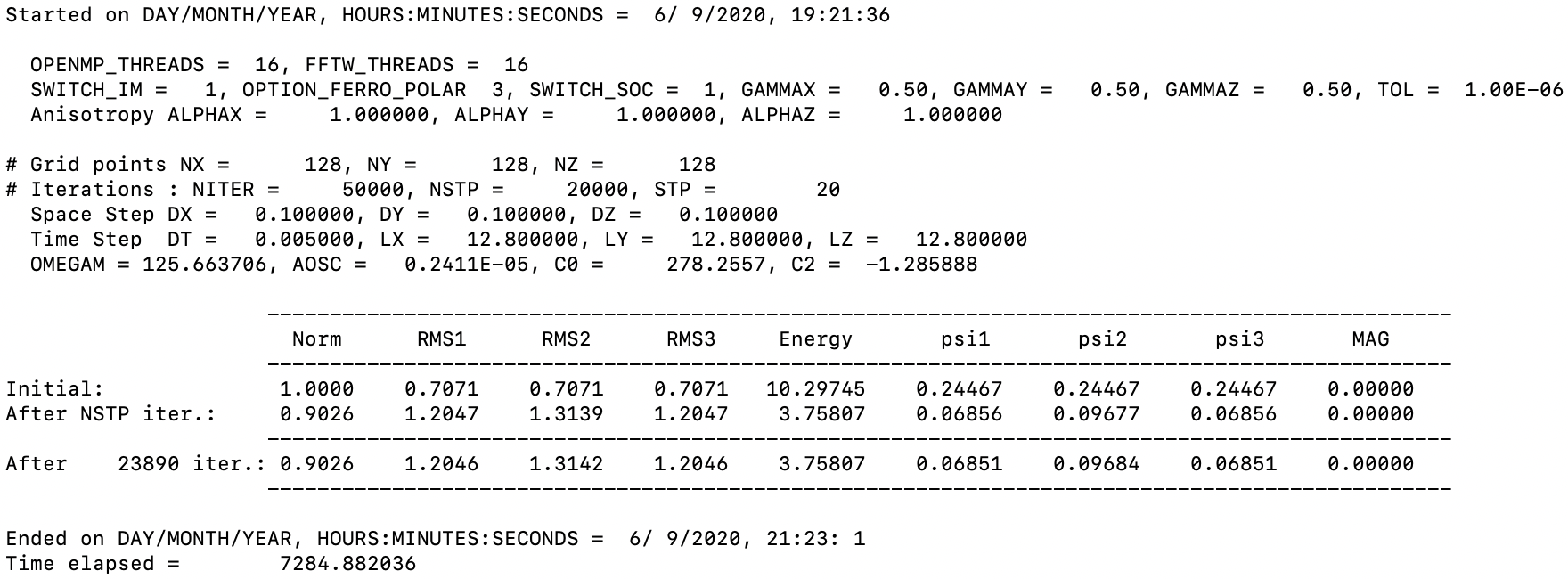}\\
 \rule{25cm}{1pt}\\{\bf (b)}
 \includegraphics[width=9.in]{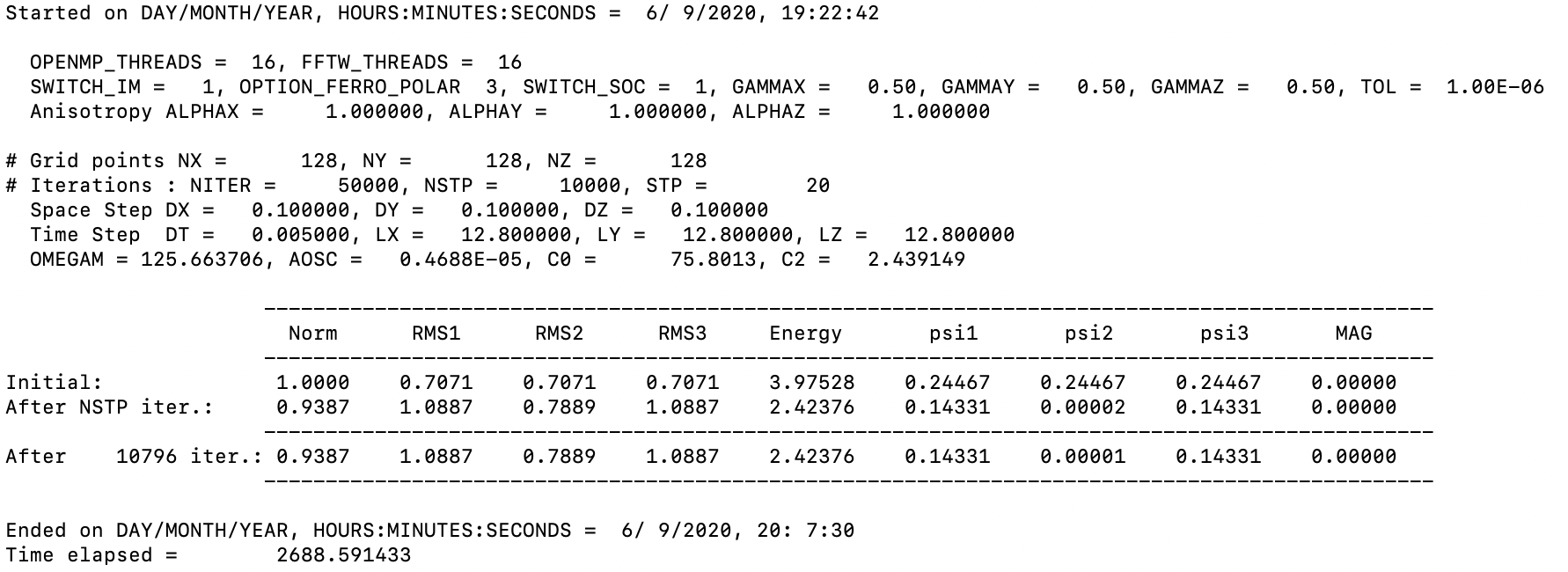}
	\caption{Contents of `file1\_im.dat' for (a) a 3D $^{87}$Rb and (b)
	a 3D $^{23}$Na condensates with isotropic SO coupling obtained with
`imretime\_spin1\_3D.f90'.}
\end{figure}